\documentclass[onecolumn,pra,aps,showpacs,nofootinbib,notitlepage]{revtex4-1}

\usepackage{graphicx}
\usepackage{subfigure}
\usepackage{dcolumn}
\usepackage{bbm}
\usepackage{color,epstopdf}
\usepackage{amscd}
\usepackage{amsfonts}
\usepackage{amsmath}
\usepackage{amssymb}

\newcommand{\mathd}{\mathrm{d}}

\newcommand{\mom}{Q}
\newcommand{\newp}{v_\eta}
\newcommand{\xz}{\alpha}
\newcommand{\pz}{\beta}
\newcommand{\sig}{\gamma}
\newcommand{\xzp}{\alpha'}
\newcommand{\om}{\Gamma}
\newcommand{\rc}{r_c}
\newcommand{\jt}{\gamma_{\text{thr}}}

\newcommand{\tmop}[1]{\ensuremath{\operatorname{#1}}}
\newcommand{\ket}[1]{\vert #1 \rangle}
\newcommand{\bra}[1]{\langle #1 \vert}
\newcommand{\scalar}[2]{\langle  #1 \vert #2 \rangle }
\newcommand{\av}[1]{\ll#1\gg}

\begin{document}
\title{Dissipative extension of the Ghirardi-Rimini-Weber model}
\author{Andrea Smirne $^{1,2}$, Bassano Vacchini $^{3,4}$, Angelo Bassi $^{1,2}$}

\affiliation{$^1$ \mbox{Dipartimento di Fisica, Universit{\`a} degli Studi di Trieste, Strada Costiera 11, I-34151 Trieste, Italy} \\
$^2$ \mbox{Istituto Nazionale di Fisica Nucleare, Sezione di Trieste, Via Valerio 2, I-34127 Trieste, Italy}\\
$^3$ \mbox{Dipartimento di Fisica, Universit{\`a} degli Studi di Milano, Via Celoria 16, I-20133 Milan, Italy}\\
$^4$ \mbox{Istituto Nazionale di Fisica Nucleare, Sezione di Milano, Via Celoria 16, I-20133 Milan, Italy}
}

\begin{abstract}

In this paper we present an extension of the Ghirardi-Rimini-Weber model for the spontaneous collapse of the wavefunction.
Through the inclusion of dissipation, we avoid the divergence of the energy on the long time scale, which
affects the original model. In particular, we define new jump operators, which depend on the momentum
of the system and lead to an exponential relaxation of the energy to a finite value. The finite asymptotic energy is naturally associated to
a collapse noise with a finite temperature, which is a basic realistic feature of our extended model.
Remarkably, even in the presence of a low temperature noise, the collapse model is effective.
The action of the new jump operators still localizes the wavefunction and the relevance of the localization increases with the size of the system,
according to the so-called amplification mechanism, which guarantees
a unified description of the evolution of microscopic and macroscopic systems.
We study in detail the features of our model, at the level
of both the trajectories in the Hilbert space and the master equation for the average state of the system.
In addition, we show that the dissipative Ghirardi-Rimini-Weber model, as well as the original one, can
be fully characterized in a compact way by means of a proper stochastic differential equation.

\end{abstract}


\maketitle

\section{Introduction}

Collapse models were formulated to describe in a unified framework the behavior of
microscopic systems, as accounted for by quantum mechanics,
and the emergence of the objective macroscopic world,
described by classical mechanics.
After the pioneering works by Pearle \cite{Pearle1976}, the first consistent collapse model
was put forward by Ghirardi, Rimini and Weber (GRW) \cite{Ghirardi1986}; see \cite{Bassi2003,Bassi2013}
for more details and a list of references about the historical development of collapse models.

The crucial feature of the GRW model is that the wavefunction associated with the state of a physical system undergoes sudden and random localization processes.
The latter do not practically affect microscopic systems, while they become relevant already on very short time scales
for macroscopic systems, by virtue of the amplification mechanism. 
The localization processes prevent macroscopic systems from being in a superposition of states centered
around macroscopically distinct positions. In addition, since any measurement process
consists in an interaction between a microscopic system and a macroscopic measurement apparatus,
the localization processes  give a dynamical
explanation of the collapse of the wavefunction
to one of the eigenstates of the measured observable, without the need of introducing an ad-hoc reduction postulate \cite{Bassi2007b}. The non-linear
and stochastic nature of the reduction postulate in standard quantum mechanics is replaced
within collapse models by a modification of the Schr{\"o}dinger equation which includes
proper non-linear and stochastic terms.

One of the main advantages of collapse models is that, besides the relevant conceptual differences
with respect to the standard theory,
they provide experimentally testable predictions which depart from those of quantum mechanics \cite{Nimmrichter2011,Romero-Isart2011}; see \cite{Bassi2013} for a
detailed list of references.
In particular, collapse models call into question the universal nature
of the superposition principle,
as they set a fundamental limit above which no physical system can exhibit position superpositions, 
except for a negligibly small time.  The investigation of collapse models 
thus plays a significant role in the experimental tests on the boundaries between the classical
and the quantum description of reality at the mesoscopic and macroscopic scale. Will it be possible
to prepare quantum superpositions for more and more complex systems with the future advances of
the experimental techniques, or is there an intrinsic limit which will prevent from this,
as predicted by collapse models?

The renewed interest in collapse models and the demand to compare their predictions 
with actual experimental data coming from very different setups has further motivated the 
formulation of more realistic
models.
The initial idea that the non-linear and stochastic modification 
of the Schr{\"o}dinger equation represents an intrinsic property of Nature, was then 
superseded by the view that the collapse of the wavefunction is induced by a physical field
filling space, which acts as a universal noise. Indeed, the precise definition of such a field
needs the formulation of a new fundamental theory going beyond standard quantum
mechanics \cite{Bassi2013}.
Hence, collapse models should be understood as phenomenological models, which
capture the main effects of the interaction with the above-mentioned noise. 
Basic physical principles set some general constraints on
the features of the admissible models. This is the case, e.g., for the absence of faster-than-light signaling \cite{Gisin1990,Bassi2013b}
or the principle of energy conservation.
An important drawback of the GRW model is the violation of the energy conservation:
the stochastic action of the noise induces larger and larger fluctuations in the momentum space,
so that the energy of the system diverges for asymptotic times, although with a rate which is very small \cite{Ghirardi1986}.
This is a common feature of the first collapse models \cite{Ghirardi1986,Diosi1989,Ghirardi1990} and it
traces back to the absence of any dissipative mechanism within the interaction
between the system and the noise \cite{Bassi2005a,Bassi2005,Vacchini2007}, as also witnessed by the
structure of the master equations which can be associated with these models.

In this paper, we modify the GRW model in order to avoid
the divergence in time of the energy, thus making an important step toward the reestablishment 
of the energy conservation within the model. 
We explicitly show that this can be achieved via the introduction
of new localization operators, without changing the other defining features of the model. This provides us with 
a more realistic collapse model, while we keep the original effectiveness and physical transparency
of the GRW model. The convergence of the energy of the system to a finite value allows us to associate the noise
with a finite temperature $T$, such that the GRW model is recovered in the high temperature limit
$T \rightarrow \infty$. 
We follow a strategy similar to that exploited for a simplified collapse model \cite{Bassi2005,Vacchini2007} and based
on the formal analogy with a Lindblad master equation \cite{Lindblad1976} including dissipation.
Besides discussing in detail the physical consequences of the extension of the model,
we also express both the original and
the generalized GRW models in terms of a stochastic differential equation.
This analysis, which relies on the framework of the stochastic differential equations of jump type in Hilbert spaces  \cite{Barchielli1991,Barchielli1994,Barchielli1995},
fills a gap with respect
to collapse models with localization continuous in time~\cite{Bassi2013}.

The paper is organized as follows. In Sec. \ref{sec:tgm}, we briefly recall
the main features of the GRW model. In Sec. \ref{sec:egm},
we introduce the dissipative GRW model, characterizing the evolution in the Hilbert space
induced by the new jump operators, in the same spirit as it was done for the original GRW model \cite{Ghirardi1986}.
In Sec. \ref{sec:sfotm}, we show how the extended GRW model, and the original one as well, can
be formulated in terms of a stochastic differential equation and we study some relevant features
of its solutions, focusing on the occurrence of the localization. In Sec. \ref{sec:me},
we derive the master equation associated with the model, which is shown to be equivalent to
an equation exploited in the description of collisional decoherence \cite{Vacchini2000, Vacchini2001,Hornberger2006}.
After studying the solution of the equation, we prove explicitly, in Sec.~\ref{sec:therm}, that it
implies dissipation and an exponential relaxation of the energy to a finite value.
In Sec. \ref{sec:ampl}, we study the amplification mechanism in the presence of dissipation: we first deal with a macroscopic rigid body and show that
its center of mass
behaves for all the practical purposes as a classical object, then we discuss
the difficulties which are unavoidably encountered when
more general situations are considered.
Finally, the conclusions and final remarks are given in Sec. \ref{sec:concl}.

\section{The GRW model}\label{sec:tgm}

\subsection{General structure of the model}\label{sec:gsotm}

Let us start by briefly recalling the main features of the collapse model introduced by Ghirardi, Rimini and Weber \cite{Ghirardi1986}. 
The GRW model can be formulated in terms of discrete jumps of the wavefunction 
which represents the state of the system taken into account. For the sake of simplicity, we consider
a particle in one dimension and we neglect spin and other internal degrees of freedom, so that
the wavefunction $\ket{\psi}$ is an element of the Hilbert space $\mathcal{L}^2(\mathbb{R})$. 
The dynamics of the particle is then characterized through the following assumptions:
\begin{itemize}
\item[1.] At random times the particle experiences a sudden jump described by
\begin{equation}\label{eq:jj}
 \ket{\psi(t)} \longrightarrow  \ket{\psi_y(t)} \equiv \frac{L_y(\widehat{X}) \ket{\psi(t)}}{\| L_y(\widehat{X}) \ket{\psi(t)} \|},
\end{equation}
where $\ket{\psi(t)}$ is the state immediately before the jump, which occurs at time $t$ and position $y \in \mathbbm{R}$, and $L_y(\hat{X}) $
is the self-adjoint contractive linear operator defined as
\begin{equation} \label{eq:locop}
L_y(\widehat{X})  = \left(\pi  \rc^2 \right)^{-1/4} e^{- (\widehat{X} - y)^2/(2  \rc^2)},
\end{equation}
with $\widehat{X}$ the position operator of the particle and $ \rc = 10^{-7} \tmop{m}$ a new parameter of the model. 
\item[2.] The overall number of jumps is distributed in time according to a Poisson process with rate $\lambda$,
which is the second new parameter of the model. The standard value of the rate is $\lambda=10^{-16} \tmop{s}^{-1}$,
while a higher rate was proposed more recently \cite{Adler2007}, $\lambda' = 2.2 \times 10^{-8 \pm 2} s^{-1}$.
\item[3.] If there is a jump at time $t$, the probability density that it takes place at the position $y$ is
\begin{equation}\label{eq:py}
p(y) = \| L_y(\widehat{X}) \ket{\psi(t)} \|^2.
\end{equation}
\item[4.] In the time interval between two consecutive jumps, the state vector evolves according 
to the usual Schr{\"o}dinger equation.

\end{itemize}

As will be shown in the following, see Sec.\ref{sec:this}, this dynamics can be equivalently formulated through a stochastic differential equation.
Moreover, the jump operators $L_y(\widehat{X})$ satisfy the relation
\begin{equation}\label{eq:normgrw}
\int \mathd y L^{\dag}_y(\widehat{X}) L_y(\widehat{X}) = \mathbbm{1},
\end{equation}
which corresponds to the normalization of the probability distribution $p(y)$ defined in Eq.(\ref{eq:py}).

\begin{figure}[!ht]
{\bf (a)}\hskip7cm{\bf (b)}\\
\includegraphics[width=.4\columnwidth]{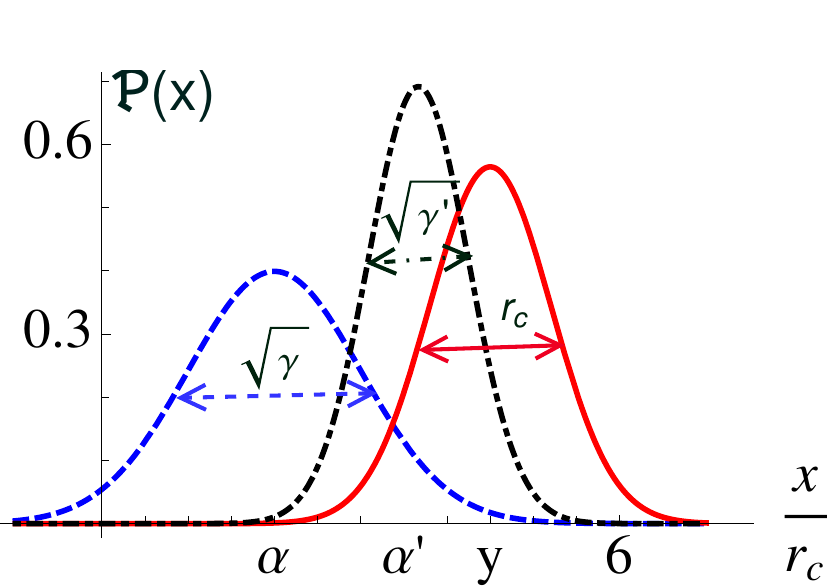}\hspace{1cm}\includegraphics[width=.43\columnwidth]{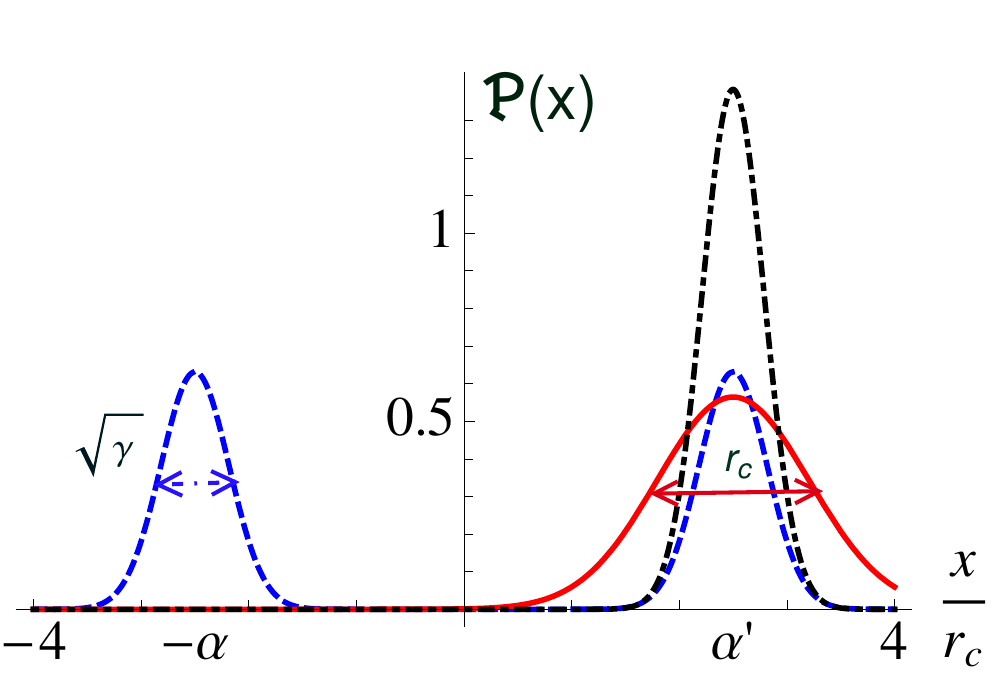}
\caption{Effect of the localization mechanism on {\bf(a)} a gaussian wavefunction $\ket{\psi} = \ket{\phi^{\xz, \pz, \sig}}$, see Eq.(\ref{eq:gaus}), 
and {\bf(b)} a superposition of gaussian wavefunctions $\ket{\psi} =\frac{1}{\sqrt{2}}\left( \ket{\phi^{\xz, 0, \sig}}+ \ket{\phi^{-\xz, 0, \sig}}\right)$, see Eq.(\ref{eq:sovr}).
The position probability distribution $\mathcal{P}(x) =|\scalar{x}{\psi}|^2$ before the localization is given by the blue dashed line,
the probability distribution after the localization  $\mathcal{P}_y(x) =|\scalar{x}{\psi_y}|^2$ is given by the black dot dashed line, while the red line represents the normalized gaussian distribution
associated with the jump operator, see Eq.(\ref{eq:locop}). The states after the localization are given by Eq.(\ref{eq:sxgrw}), {\bf (a)},
and by Eq.(\ref{eq:cmp}) with $y=\alpha$, {\bf (b)}.}
\label{fig:1}
\end{figure}

The jump operator $L_y(\widehat{X})$ describes a localization process around the position $y$ and with width $\rc$.
Consider a gaussian wavefunction $\ket{\phi^{\xz,\pz,\sig}}$, 
\begin{equation}\label{eq:gaus}
\scalar{X}{\phi^{\xz,\pz,\sig}} = C \, e^{-(X-\xz)^2/(2 \sig)} e^{i \pz  (X-\xz)/\hbar},
\end{equation}
where $X_0=\xz$ is the mean position,
$P_0 = \pz$ the mean momentum, while $\sig > 0$ determines the position variance $\Delta X^2 = \sig/2$
and the momentum variance $\Delta P^2 = \hbar^2 / (2\sig)$; $C =  \left(\pi \sig\right)^{-1/4}$ is the normalization constant.
The wavefunction $\ket{\phi_y}$ after the jump, see Eq.(\ref{eq:jj}),
is still gaussian:  apart from an irrelevant global phase, one has $\ket{\phi_y}=\ket{\phi^{\xzp,\pz,\sig'}}$,
where
\begin{eqnarray}
\xzp &=& f_{\sig} \xz +(1-f_{\sig}) y \nonumber\\
\sig'&=&\left(\frac{1}{\sig} + \frac{1}{ \rc^2}\right)^{-1},  \label{eq:sxgrw}
\end{eqnarray}
with
\begin{equation}\label{eq:fsig}
f_{\sig}\equiv \frac{\sig'}{\sig} = \left(\frac{\sig}{ \rc^2} + 1\right)^{-1}.
\end{equation}
The mean value of the momentum does not change, while the mean value of the position
is shifted toward the localization position $y$.
The position variance decreases and hence after the jump the particle
is more localized: the reciprocals of $\sig/2$
and $ \rc^2/2$ sum up and give the reciprocal of the
new variance $\sig'/2$. Importantly, the latter does not depend on the position of the localization process.  
For large gaussian wave packets,
such that $\sig \gg \rc^2$, one has $\sig' \approx  \rc^2$ and $\xz' \approx y$.
The effects of the localization process on the gaussian wavefunction are summarized in Fig.\ref{fig:1}{\bf (a)}.
In addition, the probability that the localization process occurs around $y$ is, according to Eq. (\ref{eq:py}),
\begin{equation}\label{eq:pygrw}
p(y) = \| L_y(\widehat{X})\ket{\phi^{\xz,\pz,\sig}}\|^2=  \left(\frac{f_{\sig}}{\pi  \rc^2}\right)^{1/2}e^{- (y -\xz)^2 f_{\sig}/ \rc^2}.
\end{equation}

The role of the localization operators can be further understood by means of this simple example \cite{Bassi2003}, 
which clarifies how the action of the localization operators can prevent the system
from being in a position superposition.
Consider a particle which is in the state $\ket{\varphi}$ given by the superposition of two gaussian wavefunctions with null mean momentum and the same variance, 
one centered around the position $\xz$, the other around $-\xz$:
\begin{equation}\label{eq:sovr}
\scalar{X}{\varphi} =C \left(c_+ e^{-(X-\xz)^2/(2 \sig)} + c_- e^{-(X+\xz)^2/(2 \sig)}\right),
\end{equation}
where $|c_+|^2+|c_-|^2 =1$ and $C$ is a normalization constant. 
The state after the localization around $y$,
\begin{equation}\label{eq:cmp}
\scalar{X}{\varphi_y} = C_y e^{- (X-y)^2/(2 \rc^2)}\left(c_+ e^{-(X-\xz)^2/(2 \sig)} + c_- e^{-(X+\xz)^2/(2 \sig)} \right),
\end{equation}
is still the superposition of two gaussian functions;
the normalization constant $C_y$ depends on where the localization takes place.
Now, consider a localization process around $\xz$ and assume that the distance between the two gaussians is much greater than the localization amplitude, 
while their width is much smaller than it, i.e. $\xz^2\gg \rc^2 \gg \sig$.
The previous formula directly gives  
\begin{equation}\label{eq:appst}
\scalar{X}{\varphi_{\xz}} \approx C_y \left(c_+ e^{-(X-\xz)^2\left(1/(2 \sig) + 1/(2  \rc^2)\right)} + c_- e^{-2 \xz^2/ \rc^2} e^{-(X+\xz)^2/(2\sig)} \right),
\end{equation}
so that the gaussian centered around the localization position $\xz$ is left almost unchanged,
while the gaussian centered around $-\xz$ is suppressed by a factor $e^{-2 \xz^2/ \rc^2}$. 
The localization
process practically destroys the superposition between the two gaussian wavefunctions, leading to a single gaussian state localized
around $\xz$, see Fig.\ref{fig:1}{\bf (b)}. In addition, by further exploiting $\sig \ll \rc^2$, one finds that
the probability of a localization in a neighborhood of $\pm \xz$
is given by $|c_{\pm}|^2$: the definitions in Eqs. (\ref{eq:locop}) and (\ref{eq:py})
allow to recover the usual Born's rule for the probability distributions \cite{Bassi2007b}; see also Appendix \ref{app:sovr}.
 
\subsection{Master equation associated with the model} \label{sec:amaed}
 
The GRW model is fully determined by the stochastic evolution of the wavefunction $|\psi\rangle$ previously presented. 
However, it is often convenient to deal with the dynamics of
the statistical operator which describes the state of the system averaged over all the possible trajectories built up by 
the different combinations of Schr{\"o}dinger evolutions and localization processes, see also Sec. \ref{sec:me}. 
The equation of motion of the statistical operator $\hat{\rho}(t)$, i.e., the master equation for the GRW model reads
\begin{eqnarray}\label{eq:megrw}
\frac{\mathd}{\mathd t}\hat{\rho}(t) &=& - \frac{i}{\hbar}\left[\widehat{H} \,,\, \hat{\rho}(t)\right] + \lambda\left(\int \mathd y \, L_y(\widehat{X}) \hat{\rho}(t) L_y(\widehat{X})  -\hat{\rho}(t) \right) \nonumber \\
&=&- \frac{i}{\hbar}\left[\widehat{H} \,,\, \hat{\rho}(t)\right] + \lambda\left(\left(\pi  \rc^2 \right)^{-1/2} \int \mathd y \,e^{- (\widehat{X} - y)^2/(2  \rc^2)} \hat{\rho}(t)e^{- (\widehat{X} - y)^2/(2  \rc^2)}  -\hat{\rho}(t) \right).
\end{eqnarray}
The first term describes the standard quantum evolution induced by the Hamiltonian $\widehat{H}$, while the second
term accounts for the occurrence of the localization processes.
Equation (\ref{eq:megrw}) establishes a semigroup evolution \cite{Lindblad1976} with pure decoherence in position: the off-diagonal terms
in the position representation are suppressed, and distant superpositions are suppressed faster than closer ones.

The master equation associated with the collapse model allows to investigate relevant features, such as the extension
to an $N$-particle system and then the amplification mechanism, as well as the asymptotic behavior of the energy of the system.

\subsubsection{Amplification mechanism}

Now, consider an $N$-particle system in which the localization processes occur individually for each constituent, so that the master equation
associated with the $N$-particle statistical operator $\hat{\varrho}(t)$
is simply
\begin{equation}\label{eq:megrwn}
\frac{\mathd}{\mathd t}\hat{\varrho}(t) = - \frac{i}{\hbar}\left[\widehat{H}_{\text{T}} \,,\, \hat{\varrho}(t)\right] + \sum_j \lambda_j\left(\int \mathd y \, L_y(\widehat{X}_j) \hat{\varrho}(t) L_y(\widehat{X}_j)  -\hat{\varrho}(t) \right),
\end{equation}
where $L_y(\widehat{X}_j)$ is a shorthand notation for
$\mathbbm{1}_1\otimes \ldots \mathbbm{1}_{j-1} \otimes  L_y(\widehat{X}_j) \otimes \ldots \mathbbm{1}_N$,
$\mathbbm{1}_l$ being the identity operator on the Hilbert space associated with the $l$-th particle, 
and $\widehat{X}_j$ is the position operator of the $j$-th particle, see Eq.(\ref{eq:locop}).
As we will see in Sec. \ref{sec:ampl}, the hypothesis of individual localization processes has to be considered with a certain caution.
It is useful to introduce the center of mass coordinates
through the invertible linear transformation
\begin{equation}\label{eq:rj}
\widehat{r}_j = \sum_{j'} \Lambda_{j j'} \widehat{X}_{j'},
\end{equation}
with $\Lambda_{1j}= M_j/M_{\text{T}}$, where $M_j$ is the mass of the $j$-th particle and $M_{\text{T}}= \sum^N_{j=1} M_j$
the total mass,. Accordingly
\begin{equation}\label{eq:r1}
\widehat{r}_1= \sum_j \frac{M_j}{M_{\text{T}}} \widehat{X}_j \equiv \widehat{X}_{\text{CM}}
\end{equation}
is the center-of-mass coordinate, while $(\widehat{r}_2, \ldots \widehat{r}_N)$ are the relative coordinates.
The position of the $j$-th particle can be expressed as
\begin{equation}
\widehat{X}_j = \widehat{X}_{\text{CM}} + \sum^N_{j'=2} \Lambda^{-1}_{j j'} \widehat{r}_{j'}.
\end{equation}
One can then easily prove the relation
\begin{equation}\label{eq:ampl}
 \text{Tr}_{\text{REL}}\left\{\int \mathd y \, L_y(\widehat{X}_j) \hat{\varrho}(t) L_y(\widehat{X}_j)\right\} 
 = \int \mathd y \, L_y(\widehat{X}_{\text{CM}})  \text{Tr}_{\text{REL}}\left\{\hat{\varrho}(t)\right\} L_y(\widehat{X}_{\text{CM}}),
\end{equation}
where $\text{Tr}_{\text{REL}}$ denotes the partial trace with respect to the relative degrees of freedom.
Hence, if we assume that the total Hamiltonian is the sum of a term associated with the center of mass and a term associated with the internal motion,
i.e. 
\begin{equation}\label{eq:hhh}
\widehat{H}_{\text{T}} = \widehat{H}_{\text{CM}}+\widehat{H}_{\text{REL}},
\end{equation}
we find that
the state of the center of mass, 
$
\hat{\rho}_{\text{CM}}(t) = \text{Tr}_{\text{REL}}\left\{\hat{\varrho}(t)\right\},
$
satisfies the same master equation as that in Eq.(\ref{eq:megrw}),
with the one particle Hamiltonian $\widehat{H}$ replaced by $\widehat{H}_{\text{CM}}$ 
and, most importantly, the localization rate $\lambda$ replaced by $\sum_j \lambda_j$.
This is a direct manifestation of the amplification mechanism, which, along with localization,
is the crucial feature of the GRW model. 
It explains why the model describes both microscopic and macroscopic systems. The localization rate $\lambda$ (assume $\lambda_j = \lambda$
for the sake of simplicity)
of microscopic systems is negligible and therefore the predictions of the GRW model about the wavefunction
of microscopic systems
reproduce for all practical purposes the predictions of
standard quantum mechanics. On the other hand,
if we consider a macroscopic object, the rate $\lambda_{\text{macro}} = N \lambda$, with $N$ of the order of the Avogadro's number,
induces a localization of the center of mass on very short time scales: 
the wavefunctions of macroscopic objects are almost always well-localized in space, so that
their centers of mass behave, for all the practical purposes, according to classical mechanics.

\subsubsection{Energy divergence}

A well-known drawback of the GRW model is that it predicts an infinite increase of the energy of the system. 
This energy divergence is due to larger and larger fluctuations of the momentum induced by the
interplay between the Schr{\"o}dinger evolution and the localization mechanism, as will be explicitly discussed in Sec.\ref{sec:lmaei}.
Nevertheless, the infinite increase of the energy can be inferred directly from the master equation (\ref{eq:megrw}).
In fact, this master equation predicts a linear increase of the mean value of the energy of the system 
with a rate  \cite{Ghirardi1986}
\begin{equation}
\xi = \frac{\hbar^2 \lambda}{4 M  \rc^2}.
\end{equation}
This rate of the energy increase is actually very small, $\xi = 10^{-25} eV/s$ for a nucleon, 
even if one considers the $N$-particle case \cite{Ghirardi1986}.
However, it is clear that from a fundamental point of view one would like to avoid the energy divergence and point to
a reestablishment of the energy conservation principle within the model.
This forces us to put forward a more realistic description of the interaction
between the system and the noise, also in order to test whether and how possible mechanisms excluding infinite energy increase modify
the testable predictions of the model \cite{Bassi2013}.

The use of the master equation formalism suggests a way out from the problem
of the energy divergence. Despite the deep conceptual differences between collapse models and the notion of decoherence, the
same master equation associated with the GRW model can be also derived in a specific model of collisional decoherence \cite{Vacchini2007}.
This correspondence clarifies that the origin of
the energy divergence in the GRW model can be ascribed to the lack of a dissipation mechanism, which would account
for the energy loss of the system due to the action of the noise. 
By taking into account the master equation which generalizes Eq.(\ref{eq:megrw}) to
include dissipation \cite{Vacchini2000,Vacchini2001,Hornberger2006},
we have thus been led to a possible structure of a new jump operator replacing that in Eq.(\ref{eq:locop}) and
excluding the energy divergence. 
Indeed, this was a preliminary benchmark, but the choice of the jump operator in a collapse model is subjected to further constraints, the most relevant being the induction
of localization. In addition, the definition of a collapse model in terms
of the different trajectories within the Hilbert space can be done without reference to any subsequent 
master equation or decoherence model.
For these reasons, we present our results by first postulating a new localization operator and, as a consequence, a new collapse
model and only after that we derive the corresponding master equation, see Sec.\ref{sec:me}.

\section{Extended GRW model}\label{sec:egm}

In this paper, we propose the following extension of the GRW model: the jump operators $L_y(\widehat{X})$ defined in Eq.(\ref{eq:locop})
are replaced with
\begin{equation}\label{eq:lyxp2}
L_y(\widehat{X}, \widehat{P}) = \left(\frac{\rc}{\sqrt{\pi}\hbar}+\frac{1}{2 \sqrt{\pi} M \newp}\right)^{1/2} \int \frac{\mathd \mom}{\sqrt{2 \pi \hbar}} e^{\frac{i}{\hbar} \mom (\widehat{X} - y)} 
e^{-\frac{1}{2 }\left( \left(\frac{\rc }{\hbar}+\frac{1}{2 M \newp}\right) \mom+ \frac{\widehat{P}}{M \newp}\right)^2},
\end{equation}
where $\widehat{P}$ is the momentum operator of the system, $M$ the mass of the particle and 
\begin{equation}
\newp= 10^{31} \frac{\hbar / \rc}{\text{Kg}}
\end{equation}
a new parameter of the model,
which is related with the temperature of the noise inducing the localization. This will be explicitly shown in Sec.\ref{sec:therm},
where the specific choice of $\newp$, as well as its peculiar role in the definition of $L_y(\widehat{X}, \widehat{P})$, will be discussed.
Note that the jump operators of the model are no longer self-adjoint. 
An equivalent way to express $L_y(\widehat{X}, \widehat{P})$
is given by
\begin{equation}\label{eq:lyxp}
L_y(\widehat{X}, \widehat{P}) = \left(\frac{\sqrt{\pi}\hbar}{2 M \newp} + \sqrt{\pi}\rc \right)^{-1/2} \int \mathd X \mathd P \ket{X} \bra{X}
 e^{-\frac{(X - y)^2}{2 \left(\hbar/(2 M \newp) + \rc \right)^{2}}} e^{- \frac{ i(X-y)  P}{M \newp \rc + \hbar/2 }}  \ket{P} \bra{P},
\end{equation}
by which one immediately sees how the original GRW jump operator $L_y(\widehat{X})$, see Eq.(\ref{eq:locop}),
is obtained in the limit $\newp \rightarrow \infty$.
On the other hand, the general structure provided by items 1-4 in the previous section
is left untouched. Explicitly, our collapse model can be formulated as follows:
\begin{itemize}
\item[1.] The sudden jumps are now described by
\begin{equation}\label{eq:jjd}
 \ket{\psi(t)} \longrightarrow  \ket{\psi_y(t)} \equiv \frac{L_y(\widehat{X}, \widehat{P}) \ket{\psi(t)}}{\| L_y(\widehat{X}, \widehat{P}) \ket{\psi(t)} \|}.
\end{equation}
\item[2.] The overall number of jumps is still distributed in time according to a Poisson process with rate $\lambda$.
\item[3.] If there is a jump at time $t$, the probability density that the jump takes place at the position $y$ is now given by
\begin{equation}\label{eq:pyd}
p(y) = \| L_y(\widehat{X}, \widehat{P}) \ket{\psi(t)} \|^2.
\end{equation}
\item[4.] Still, between two consecutive jumps the state vector evolves according to the Schr{\"o}dinger equation.
\end{itemize}
It is important to observe that the jump operators satisfy a normalization condition as in Eq.(\ref{eq:normgrw}); explicitly,
\begin{eqnarray}
\int \mathd y \, L^{\dag}_y(\widehat{X}, \widehat{P}) L_y(\widehat{X}, \widehat{P}) &=& 
 \left(\frac{\rc}{\sqrt{\pi}\hbar}+\frac{1}{2 \sqrt{\pi} M \newp}\right)  \int \frac{\mathd y\,  \mathd \mom \, \mathd \mom' }{2 \pi \hbar}
e^{-\frac{1}{2 }\left( \left(\frac{\rc }{\hbar}+\frac{1}{2 M \newp}\right) \mom+ \frac{\widehat{P}}{M \newp}\right)^2}
e^{-\frac{i}{\hbar} \mom (\widehat{X} - y)}  \nonumber\\
&&\times e^{\frac{i}{\hbar} \mom' (\widehat{X} - y)} 
e^{-\frac{1}{2 }\left( \left(\frac{\rc }{\hbar}+\frac{1}{2 M \newp}\right) \mom'+ \frac{\widehat{P}}{M \newp}\right)^2} \nonumber\\
&=& \left(\frac{\rc}{\sqrt{\pi}\hbar}+\frac{1}{2 \sqrt{\pi} M \newp}\right)  \int \,  \mathd \mom \, \mathd \mom' 
e^{-\frac{1}{2 }\left( \left(\frac{\rc }{\hbar}+\frac{1}{2 M \newp}\right) \mom+ \frac{\widehat{P}}{M \newp}\right)^2}
e^{-\frac{i}{\hbar} \mom \widehat{X} }  \nonumber\\
&&\times e^{\frac{i}{\hbar} \mom' \widehat{X} } 
e^{-\frac{1}{2 }\left( \left(\frac{\rc }{\hbar}+\frac{1}{2 M \newp}\right) \mom'+ \frac{\widehat{P}}{M \newp}\right)^2}
\delta(\mom-\mom') \nonumber\\
&=&
 \left(\frac{\rc}{\sqrt{\pi}\hbar}+\frac{1}{2 \sqrt{\pi} M \newp}\right)  \int \,  \mathd \mom 
e^{-\left( \left(\frac{\rc }{\hbar}+\frac{1}{2 M \newp}\right) \mom+ \frac{\widehat{P}}{M \newp}\right)^2}
= \mathbbm{1}. \label{eq:norml}
\end{eqnarray}
This property guarantees that the probability distribution $p(y)$ associated with
the localization position is properly normalized, see Eq.(\ref{eq:pyd}), and 
its role within the model will be further discussed at the end of Sec. \ref{sec:edsme}. 
As will be shown extensively in the following, the replacement of 
 $L_y(\widehat{X})$ with $L_y(\widehat{X}, \widehat{P})$ preserves
all the desired features of the resulting collapse model, while
the dependence on the momentum operator $\widehat{P}$ prevents the infinite energy increase and thus leads to a more realistic description of the action of the noise.

The role of the jump operator $L_y(\widehat{X}, \widehat{P})$ is illustrated directly by evaluating its action on 
a gaussian wavefunction $\ket{\phi^{\xz,\pz,\sig}}$, see Eq.(\ref{eq:gaus}).
The gaussian structure of the wavefunction is preserved and, specifically, one has
that the state after the jump, see Eq.(\ref{eq:jjd}), is 
$
\ket{\phi_y}= \ket{\phi^{\xzp, \pz',\sig'}}$,
with
\begin{eqnarray}
\xzp &=&  g_{\sig} \xz +(1-g_{\sig}) y \nonumber\\
\pz' &=& \pz \frac{1 -k}{1+k} \nonumber\\
\sig' &=& \left(\frac{(1-k)^2}{\sig (1+k)^2}+\frac{1}{ \rc^2(1+k)^2}\right)^{-1}, \label{eq:xps}
\end{eqnarray}
where we introduced the adimensional quantity
\begin{eqnarray}
 k &\equiv& \frac{\hbar}{2 M \newp  \rc} = 5\times10^{-32} \frac{\text{Kg}}{M},  \label{eq:kg}
\end{eqnarray} 
which will be crucial in the following analysis,
as well as 
\begin{eqnarray}
 g_{\sig} &\equiv&  \frac{(1-k)\sig'}{(1+k)\sig} = \left(\frac{\sig}{ \rc^2(1-k^2)} + \frac{1-k}{1+k} \right)^{-1}. \label{eq:kg2}
\end{eqnarray}
The jump shifts the mean value of the position toward $y$, and, now, 
it also damps the mean value of the momentum.
The variance $\sig'$ after the jump does not depend on where the jump takes place and
it is given by the reciprocal of the sum of the reciprocals of $ \rc^2(1+k)^2$ and $\sig(1+k)^2/(1-k)^2$. The contribution due to the
$\widehat{P}$-dependent term causes a slight increase of the wave-function width, which partially counterbalances the decrease
due to the usual GRW contribution. 
For gaussian wavefunctions such that $\sig$ is smaller than the threshold value
\begin{equation}\label{eq:soglia}
\jt \equiv 4 k \rc^2 = \left(2* 10^{-45} \frac{\text{Kg}}{m}\right) \text{m}^{2}
\end{equation}
the jump induced by $L_y(\widehat{X}, \widehat{P})$ will increase overall the position variance.
Contrary to the original GRW model, the repeated action of the jump operators does not induce
an unlimited contraction of the wavefunction:  there is a lower threshold under which the jump processes cease to be localization processes.
In realistic situations, this threshold value is not reached by the evolution, see Eq.(\ref{eq:asvar}) and the following discussion.
However, $\jt$ plays a crucial role in fixing the asymptotic finite value
of the energy, see Sec. \ref{sec:therm}.
Let us note that, as in the original GRW model, the wavefuction after the localization
process is non-vanishing over the whole space. This is the so-called 'problem of the tails'
in collapse models \cite{Lewis1997,Clifton1999,Bassi1999,Bassi2003,Wallace2008} and, indeed, the introduction of dissipation leaves it unaltered.

The probability density for a jump to take place at the position $y$
is, see Eq.(\ref{eq:pyd}),
\begin{equation}\label{eq:prdi}
p(y) = \| L_y(\widehat{X}, \widehat{P})\ket{\phi^{\xz,\pz,\sig}}\|^2=  \left(\frac{g_{\sig}}{\pi (1-k^2)  \rc^2}\right)^{1/2}e^{-\frac{(y -\xz)^2 g_{\sig}}{(1-k^2) \rc^2}}.
\end{equation}
Finally, by taking into account the action of $L_y(\widehat{X}, \widehat{P})$ on a superposition of two gaussian wavefunctions as in Eq.(\ref{eq:sovr}),
with $\xz^2\gg \rc^2\gg\sig$, one can show along the same lines as for the original GRW model, see Appendix \ref{app:sovr}, that also in this case the localization process destroys 
the superposition and selects
a single gaussian wavefunction, see Fig.\ref{fig:1} {\bf (b)}, with a probability which corresponds to the usual Born's rule.

\section{Trajectories in the Hilbert space}\label{sec:sfotm}
In the previous section, we have extended the GRW model by introducing
new jump operators, while leaving the general structure of the collapse model untouched. 
The dynamics of the wavefunction consists in a unitary evolution interrupted
by sudden discontinuous transformations (jumps) at random and separated times. 
The stochastic differential equations which usually define collapse models are 
governed by Wiener processes  \cite{Ghirardi1990,Bassi2013},
so that they do not supply the piecewise deterministic evolution now recalled.
However, we will show in this section how also the generalized GRW model can be 
formulated by postulating a stochastic differential equation.
The latter determines the trajectories in the Hilbert space of the system through a random field,
i.e. a family of stochastic processes, one for each point of space $y \in \mathbbm{R}$. 
As for the other collapse models, this equation has to be understood as a phenomenological equation,
whose fundamental motivation has to be looked for by some underlying theory beyond standard quantum mechanics \cite{Bassi2013}.
In the proper limit, we will also get a stochastic differential equation for the original GRW model.

The reader is referred to \cite{Barchielli1991,Barchielli1994,Barchielli1995}
for further details and a rigorous treatment of the stochastic differential equations with jumps
in Hilbert spaces. 

\subsection{Stochastic differential equation}\label{sec:this}
First, let us introduce a family of stochastic processes $\left\{N_y(t)\right\}_{y\in\mathbb{R}}$
such that the counting process $N_y(t) \mathd y$ counts
the jumps taking place at a position within $y$ and $y+\mathd y$.
The stochastic processes are defined on a common probability space $(\Omega, \mathcal{F}, \mathbbm{P})$ and $\mathbbm{E}[\cdot]$
indicates the statistical mean with respect to the probability $\mathbbm{P}$.
Furthermore, we denote as $\omega_t  = (t_1, y_1; t_2, y_2; \ldots t_m, y_m)$ 
a generic sequence of instants and positions in which the jumps occur up to time $t$;
indeed, this corresponds to specifying the trajectories of the counting processes up to time $t$.
We assume that the processes $\left\{N_y(t)\right\}_{y\in\mathbb{R}}$ are independent and satisfy
\begin{eqnarray}
\mathd N_{y}(t) \mathd t &=& 0 \label{eq:edn1} \\
\mathd N_{y'}(t) \mathd N_{y}(t) &=& \delta (y'-y) \mathd N_y(t) \label{eq:edn2}\\
\mathbbm{E}[\mathd N_y(t) | \omega_t] &=&\lambda \| L_y(\widehat{X}, \widehat{P}) \ket{\psi(t)}\|^2 \mathd t, \label{eq:edn3}
\end{eqnarray}
where $\mathd N_y(t) = N_y(t+\mathd t) - N_y(t)$ is the increment of  $ N_y(t)$ in a time $\mathd t$.
We introduced the short-hand notation $ \ket{\psi(t)} \equiv  \ket{\psi(\omega_t)}$ to indicate the state
of the system depending on the trajectory up to time $t$: 
the wavefunction is itself a stochastic process, which has values in the Hilbert space associated with the
system and is determined by the sequences of jumps, i.e. by the trajectories of the counting processes.
Equations (\ref{eq:edn1}) and (\ref{eq:edn2})
tell us that the probability of one count in a time interval $\mathd t$ is of order $\mathd t$,
while the probability of more than one count is of higher order \cite{Barchielli1994}.
Equation (\ref{eq:edn3}) yields the expected value of the
increment of the counting
processes  \emph{conditioned} on the occurrence of the sequence of jumps $\omega_t$ up to time $t$ \cite{Barchielli1991,Barchielli1994}.
This conditional expected value depends both on the (stochastic) wavefunction at time $t$
and on the (deterministic) jump operator  $L_y(\widehat{X}, \widehat{P})$.
Finally, the wavefunction $\ket{\psi(t)}$  is fixed by the following non-linear stochastic differential equation:
\begin{equation}\label{eq:edsnl}
\mathd \ket{\psi(t)} =- \frac{i}{\hbar} \widehat{H}  \ket{\psi(t)} \mathd t + \int \mathd y \left(\frac{L_y(\widehat{X}, \widehat{P})}{\|L_y(\widehat{X}, \widehat{P}) \ket{\psi(t)}\|} - \mathbbm{1}\right) \ket{\psi(t)} \mathd N_y(t).
\end{equation}
The evolution of 
the wavefunction in a time interval $\mathd t$ has a deterministic contribution
due to the Hamiltonian $\widehat{H}$ and a stochastic contribution due 
the jumps described by Eq.(\ref{eq:jjd});
a jump around the position $y$ corresponds to a non-zero increment of the counting process $N_y(t) \mathd y$.
The solution of Eq.(\ref{eq:edsnl}) can be represented straightforwardly: 
given the sequence of jumps $\omega_t  = (t_1, y_1; t_2, y_2; \ldots; t_m, y_m)$ and the initial condition $\ket{\psi(t_0)} \equiv \ket{\psi}_0$,
the corresponding trajectory in the Hilbert space is
\begin{equation} \label{eq:sol}
\ket{\psi(t)} = \frac{1}{C(\omega_t)} e^{- i \widehat{H} (t-t_m) / \hbar} L_{y_m}(\widehat{X}, \widehat{P}) \ldots  e^{- i \widehat{H} (t_2-t_1) / \hbar} L_{y_1}(\widehat{X}, \widehat{P}) 
 e^{- i \widehat{H} (t_1-t_0) / \hbar} \ket{\psi}_0,
\end{equation}
where $C(\omega_t) \equiv \|e^{- i \widehat{H} (t-t_m) / \hbar} L_{y_m}(\widehat{X}, \widehat{P}) \ldots  e^{- i \widehat{H} (t_2-t_1) / \hbar} L_{y_1}(\widehat{X}, \widehat{P}) 
e^{- i \widehat{H} (t_1-t_0) / \hbar} \ket{\psi}_0 \|^{1/2}$ is the normalization factor.
This equation formally characterizes all the possible evolutions of the system's state within our model.
The deterministic evolution of the wavefunction induced by the group of unitary operators $U(t) = e^{- i \widehat{H} t / \hbar}$
is interrupted by the jumps described by the operators $L_{y}(\widehat{X}, \widehat{P})$;
the dynamics introduced in Sec. \ref{sec:egm} is then recovered,
compare with items $1$ and $4$.

Furthermore, all the other features of the collapse model can be retrieved by the properties of the stochastic processes in Eqs.(\ref{eq:edn1})-(\ref{eq:edn3}). 
Let us say that the system is in the state $\ket{\psi(t)}$ at time $t$ and recall
that the probability 
of more than one jump in a time interval $\mathd t$ is negligible.
Hence, the probability density $p(y, t | \psi(t)) \mathd t$ of a jump at the position $y$ and a time between $t$ and $t+\mathd t$ is simply given
by the conditional expectation of the increment of the corresponding process $N_y(t)$, i.e., \cite{Barchielli1991}
\begin{equation}\label{eq:pro}
p(y, t | \psi(t)) =  \mathbbm{E}[\mathd N_y(t) | \omega_t] =\lambda \| L_y(\widehat{X}, \widehat{P}) \ket{\psi(t)}\|^2 \mathd t.
\end{equation} 
Now, as the jump operators satisfy the normalization condition in Eq.(\ref{eq:norml}),
the probability $p(t |\psi(t))$ to have a jump within $t$ and $t + \mathd t$ at any position is simply
\begin{equation}
p(t |\psi(t)) = \int \mathd y \, p(y, t |\psi(t)) = \lambda \int \mathd y \bra{\psi(t)}  L^{\dag}_y(\widehat{X}, \widehat{P}) L_y(\widehat{X}, \widehat{P}) \ket{\psi(t)} \mathd t =   \lambda \mathd t,
\end{equation}
i.e. the overall jump rate does not depend on the state of the system and is given by $\lambda$, according to item $2.$ 
As a matter of fact, this corresponds to the rate of the Poisson process
\begin{equation}\label{eq:np}
N(t) = \int \mathd y N_y(t),
\end{equation}
which counts the total number of jumps up to time $t$.
Finally, the probability density that, if there is a jump at a time between $t$ and $t+\mathd t$,
it takes place at the position $y$ is $ p(y, t |\psi(t))/  p(t |\psi(t)) =  \| L_y(\widehat{X}, \widehat{P}) \ket{\psi(t)}\|^2$,
so that Eq.(\ref{eq:pyd}) and item $3$ are recovered.

The results of this paragraph apply to the original GRW model in the limit $\newp\rightarrow \infty$, i.e. 
$L_y(\widehat{X}, \widehat{P})\rightarrow  L_y(\widehat{X})$.
For example,
we can associate the GRW model with the non-linear stochastic differential equation
\begin{equation}\label{eq:edsnlgrw}
\mathd \ket{\psi(t)} =- \frac{i}{\hbar} \widehat{H}  \ket{\psi(t)} \mathd t + \int \mathd y \left(\frac{e^{- (\widehat{X} - y)^2/(2  \rc^2)}}{\|e^{- (\widehat{X} - y)^2/(2  \rc^2)} \ket{\psi(t)}\|} - \mathbbm{1}\right) \ket{\psi(t)} \mathd N_y(t).
\end{equation}
By following  \cite{Barchielli1991,Barchielli1994,Barchielli1995}, one can also introduce a linear equation equivalent to Eq.(\ref{eq:edsnl}) after a proper change of probability
on the measurable space $(\Omega, \mathcal{F})$.

\subsection{Position and momentum localization}\label{sec:paml}

In Appendix \ref{sec:gs}, we study in detail the gaussian solutions of the stochastic differential equation (\ref{eq:edsnl}).
Here, we focus on the evolution of the position variance, thus confirming the
effectiveness of the localization mechanism ruling the collapse model.
We further characterize the finite values of the position and the momentum variances 
in the asymptotic time limit.

\subsubsection*{Localization of gaussian wavefunctions}
Given a gaussian solution of Eq.(\ref{eq:edsnl}), i.e. $ \ket{\phi^{\xz_t,\pz_t,\sig_t}}$  as in Eq.(\ref{eq:gaus}),
with $\xz_t, \sig_t \in \mathbbm{C}, \pz_t \in \mathbbm{R}$ and the normalization $C$
in Eq.(\ref{eq:cc}),
the position variance $(\Delta_{\phi_t} X)^2$ is defined as, compare with Eq.(\ref{eq:apvar}),
\begin{equation}\label{eq:apvar}
(\Delta_{\phi_t} X)^2  =  \bra{\phi^{\xz_t,\pz_t,\sig_t}} \widehat{X}^2\ket{ \phi^{\xz_t,\pz_t,\sig_t}}-(\bra{\phi^{\xz_t,\pz_t,\sig_t}} \widehat{X}\ket{ \phi^{\xz_t,\pz_t,\sig_t}})^2.
\end{equation}
As shown in Appendix \ref{sec:gs},  $(\Delta_{\phi_t} X)^2$ depends on the instants of the jumps, but not on their position, 
and hence it is a function of the trajectories $\varpi_t=(t_1,t_2, \ldots t_m)$ of the Poisson process $N(t)$.
To illustrate in a compact way the evolution of the position variance,
we deal with its statistical mean.
For a Poisson process with rate $\lambda$, the probability that there is one count between $t_1$ and $t_1+\mathd t_1$, $\ldots$, 
one count between $t_m$ and $t_m+\mathd t_m$ and no other counts up to time $t$ is \cite{Barchielli1994} $\lambda^{m} e^{- \lambda t} \mathd t_1 \ldots \mathd t_m$:
the probability density associated with $\varpi_t$ only depends on the overall time $t$ and number of jumps $m$.
Thus, the expected value of the position variance reads
\begin{equation}\label{eq:pv}
\mathbbm{E}[(\Delta_{\phi_t} X)^2] = \sum^{\infty}_{m=0} \lambda^m e^{- \lambda t} \int^t_0 \mathd t_m \ldots \int^{t_2}_0 \mathd t_1
 \frac{\left| \mathcal{G}_m\left(\mathcal{G}_{m-1}\left(\ldots \mathcal{G}_1(\gamma)\right)\right) +\frac{i \hbar (t-t_m)}{M} \right|^2}{2 \text{Re}[ \mathcal{G}_m\left(\mathcal{G}_{m-1}\left(\ldots \mathcal{G}_1(\gamma)\right)\right) +\frac{i \hbar (t-t_m)}{M}]},
\end{equation}
where $\mathcal{G}_j(x)$ is defined in Eq.(\ref{eq:f}), and the integrand expresses the position variance at time $t$ on the trajectory $\varpi_t=(t_1,t_2, \ldots t_m)$, see Eqs. (\ref{eq:fff})
and (\ref{eq:varx}).
\begin{figure}[!ht]
{\bf (a)}\hskip7cm{\bf (b)}\\
\includegraphics[width=.39\columnwidth]{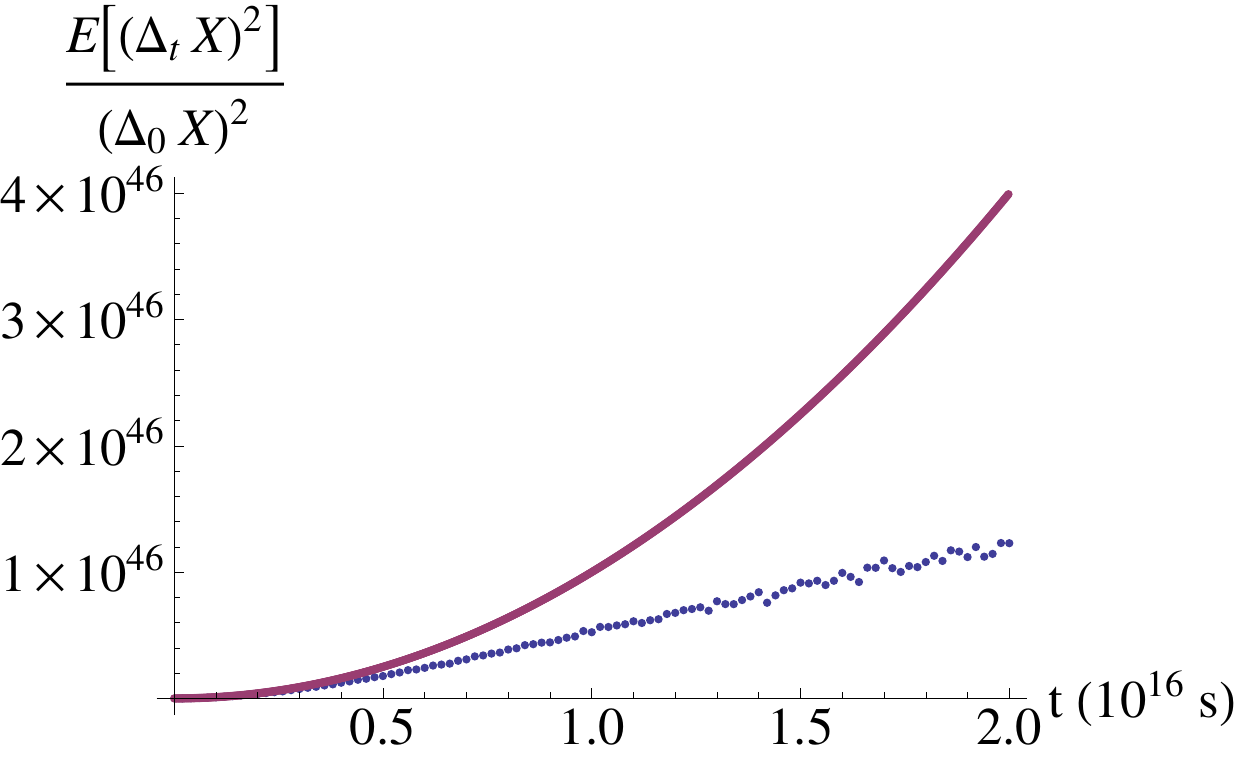}\hspace{0.5cm}\includegraphics[width=.39\columnwidth]{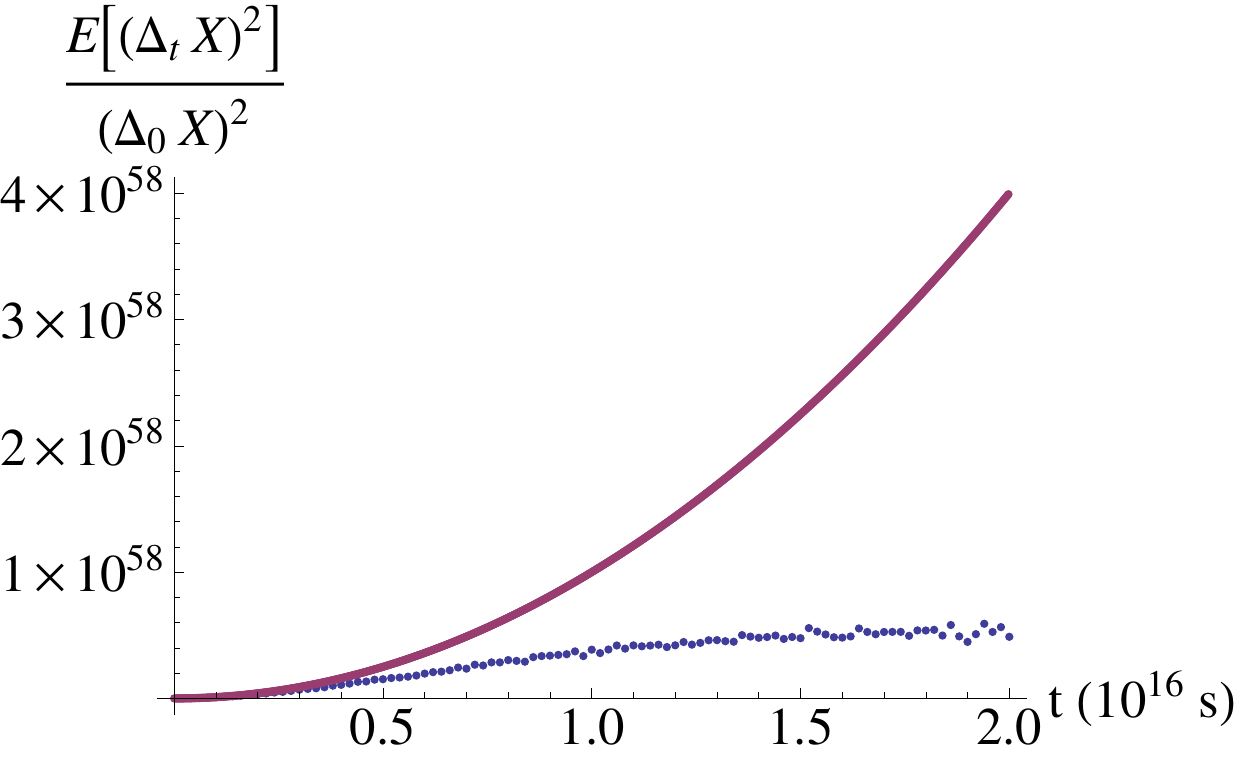}\\
\vspace{0.5cm}
{\bf (c)}\hskip7cm{\bf (d)}\\
\includegraphics[width=.39\columnwidth]{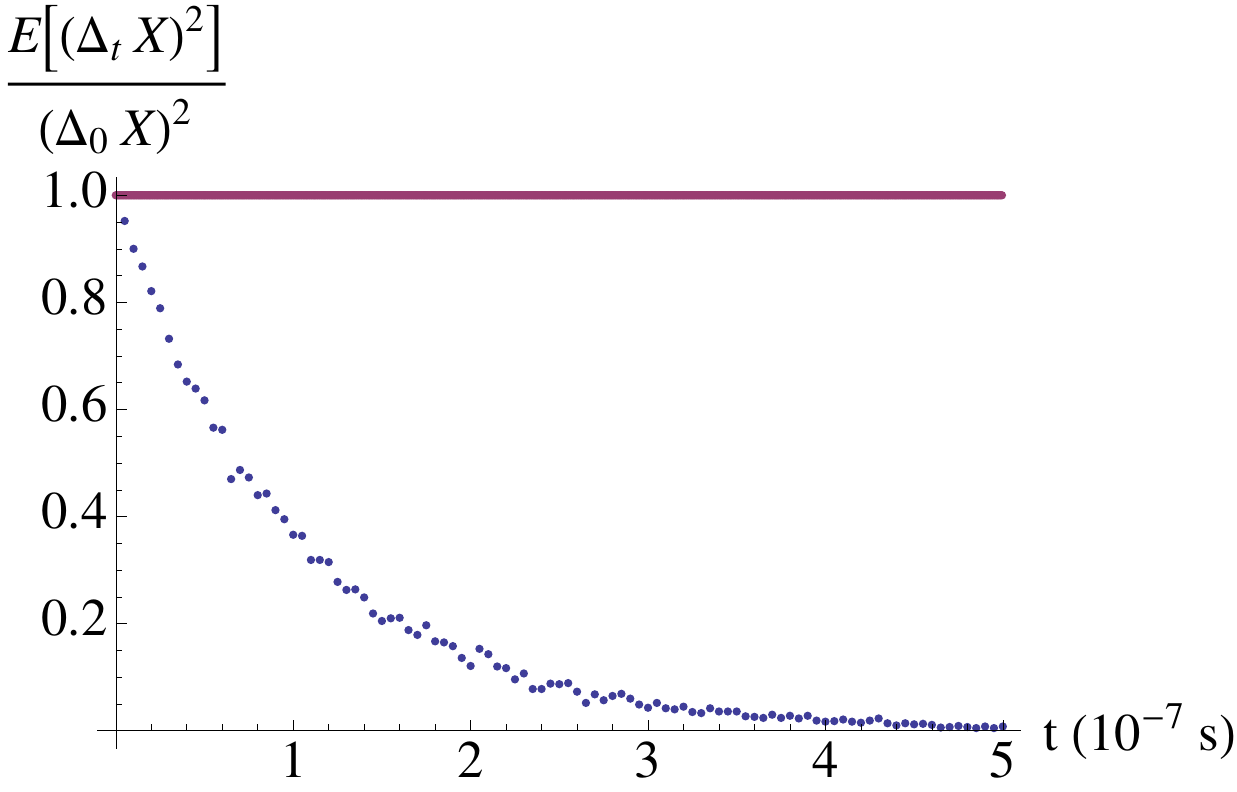}\hspace{0.5cm}\includegraphics[width=.39\columnwidth]{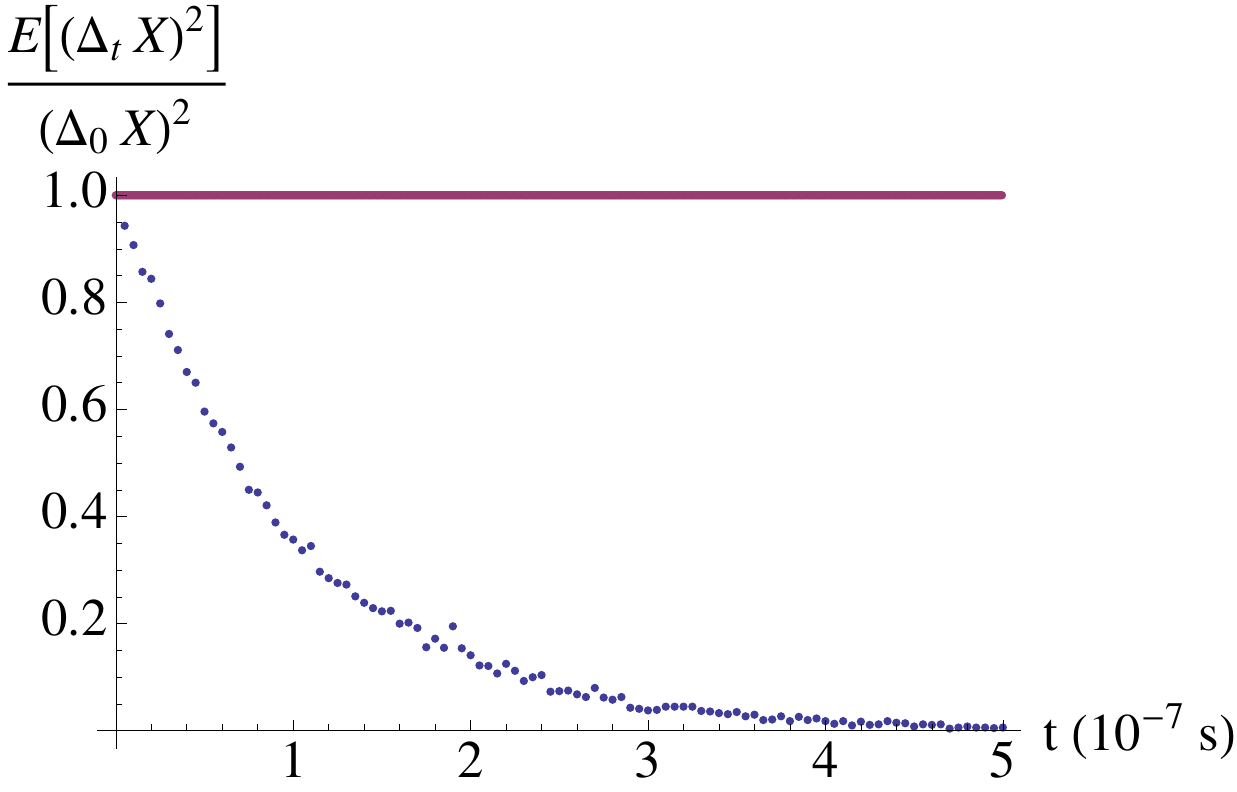}
\caption{Expected value of the position variance, see Eq.(\ref{eq:pv}),  
as a function of time (blue dots), compared with the deterministic unitary evolution (red line);
the statistical mean is obtained over a sample
of $10^5$ trajectories. ({\bf{a}}, {\bf{b}})  Microscopic system: $M = 10^{-27} Kg$ and jump rate given by $\lambda = 10^{-16} s^{-1}$, with
$\gamma = \rc^2$ ({\bf{a}}) and $\gamma = 10^{-6} \rc^2$ ({\bf{b}}).
({\bf{c}}, {\bf{d}})  Macroscopic system: $M = 10^{-3} Kg$ and jump rate given by $\lambda_{\text{macro}} = N \lambda = 10^{7} s^{-1}$, with
$\gamma = 10^6 \rc^2 $ ({\bf{c}}) and $\gamma = 10^{12}\rc^2$ ({\bf{d}}). The initial variance is $(\Delta_{\phi_0} X)^2 = \sig/2$.}
\label{fig:loc}
\end{figure}In Fig. \ref{fig:loc}, we can observe the evolution of $\mathbbm{E}[(\Delta_{\phi_t} X)^2]$ for different values of the 
initial variance $\sig/2$ and for both the microscopic and the macroscopic regime.
The former refers to the evolution of a single particle with a mass of the order of the nucleon mass $M=10^{-27} kg$,
while the latter describes the evolution of the center of mass of a system composed by an Avogadro's number $N$
of particles. As will be shown in Sec. \ref{sec:ampl},
we can apply our collapse model to an $N$-particle system
by simply replacing the jump rate $\lambda$ with $\lambda_{\text{macro}} = N \lambda$ and referring $M$ to the total mass of the system,
at least as long as a rigid body is considered. In the microscopic
regime, Fig. \ref{fig:loc}.({\bf{a}}) and ({\bf{b}}), the evolution of the expected value of the position variance
strictly follows the deterministic unitary evolution up to very long time scales
and then saturates to a finite value, see the next paragraph. 
The jump rate is $10^{-16} s^{-1}$ and therefore the probability to have a jump will
be negligible up to, say, $10^{15} s$: the action of the noise does not induce any observable localization process
on the microscopic systems.
On the other hand, when macroscopic systems are taken into account, Fig. \ref{fig:loc}.({\bf{c}}) and ({\bf{d}}),
the evolution described by Eq.(\ref{eq:pv}) strongly departs from the unitary one
from the very beginning of the dynamics. The repeated occurrence of the jumps
rapidly reduces the position spread of the wavefunction,
so that the localization mechanism is clearly manifested.
The time scale of the wavefunction localization is $\lambda_{\text{macro}}^{-1}$
and it is the same as that for the GRW model. The total rate of events in our extended GRW model
is in fact the same as in the original one, compare items 2 of Secs. \ref{sec:egm} and \ref{sec:gsotm}.
We conclude that the modification of the jump operators put forward with Eq.(\ref{eq:lyxp2}) does not
introduce any significant change in the localization mechanism compared to the original GRW model,
as also shown by the results of the next paragraph.

\subsubsection*{Asymptotic values of position and momentum variances}\label{sec:lmaei}

The trajectories of the model are made up of a sequence of
deterministic unitary evolutions and random jumps. These two
transformations have opposite effects on the wavefunction, as long as the position 
variance is concerned. The free evolution induces a spread of the position variance, which is the faster
the narrower the wavefunction. On the contrary, the jumps shrink the wavefunction, at least as long as $\sig > \sig_{\text{thr}}$, see
Eq.(\ref{eq:soglia}).
At some point of the evolution the two opposite effects balance each other
and thus the position variance reaches a finite and non-zero equilibrium value \cite{Ghirardi1986,Bassi1999}. 
As shown in Appendix \ref{app:axp},
the asymptotic value of the position variance can be evaluated via the relation
\begin{equation}\label{eq:asvar}
(\Delta_{\phi} X)_{\text{as}}^2 = \frac{\rc^2(1+k)^2 }{1 +  \sqrt{\frac{1}{2}\left(\chi- \jt^2 /\epsilon^2  + 1 \right)}},
\end{equation}
with
\begin{equation}
\chi = \sqrt{ \jt^4/\epsilon^4 +2(\jt^2- 8 \jt \rc^2(1+k)^2  + 8 \rc^4(1+k)^4)/ \epsilon^2 + 1}.
\end{equation}
This asymptotic value is in general
much higher than the value $\jt/2$, which would correspond
to the threshold in Eq.(\ref{eq:soglia}).
For a macroscopic system, with $M = 10^{-3} \text{Kg}$, one has 
$\jt/2 = 10^{-42} \text{m}^2$, while $ (\Delta_{\phi} X)_{\text{as}}^2 \approx 7 \times  10^{-26} \text{m}^2$, 
which is also in agreement with the estimate given in \cite{Ghirardi1986}
for $k=0$. As a matter of fact, due to the specific choice of $k$, the threshold value is very small, so that the free evolution
and the jumps balance each other before the spread of the wavefunction can reach it.

Analogously, see Appendix \ref{app:axp}, the asymptotic value of the momentum variance is given by
\begin{equation}\label{eq:deltap}
(\Delta_{\phi} P)_{\text{as}}^2 = \frac{\hbar^2 }{ \jt + \epsilon \sqrt{\frac{1}{2}\left(\chi + \jt^2/\epsilon^2 - 1 \right)}}.
\end{equation}
For a macroscopic system with $M=10^{-3} \text{Kg}$, one gets $(\Delta_{\phi} P)_{\text{as}}^2 \approx 7 \times 10^{-43} \text{Kg}\,\text{m}\,\text{s}^{-1}$,
still perfectly compatible with the value for $k=0$ \cite{Ghirardi1986}, so that $(\Delta_{\phi} X)_{\text{as}}^2(\Delta_{\phi} P)_{\text{as}}^2$
is approximately twice the minimum value allowed by the uncertainty relation.

It is worth noting that, as the position variance, also the momentum
variance reaches a finite asymptotic value, both for $k \neq 0$ and for $k = 0$, which naturally leads to the following remark.
The reason for the energy divergence in the original GRW model is  quite a subtle one. It is often understood by saying that
the jump operator in Eq.(\ref{eq:locop}) induces an indefinite contraction of the width of the wavefunction,
so that $(\Delta_{\phi_t} X)^2 \rightarrow 0$ and therefore, in accordance with the uncertainty relation, $(\Delta_{\phi_t} P)^2 \rightarrow \infty$,
implying the divergence of the energy. However, it is clear
how this picture is not the end of the story and it is, to some extent, misleading. 
A crucial role here is played by the Schr{\"o}dinger evolution between the jumps. 
As now recalled, the balance between unitary evolution and jumps implies a finite
asymptotic value of the momentum variance, also for $k=0$.
This means that the energy divergence in the GRW model is actually due to fluctuations
of the mean value of the momentum, as $\langle H \rangle_t = ((\Delta_{\phi_t} P)^2 + \langle P\rangle^2_t)/(2M)$. To be more explicit, 
consider an initial gaussian wavefunction, $\ket{\psi}_0 = \ket{\phi^{\xz,\pz,\sig}}$.
The unitary evolution up to the first jump at time $t_1$ does not modify the mean value of the momentum $\langle P \rangle_{t_1} = \pz $
and shifts the mean value of
the position as $\langle X \rangle_{t_1} = \xz + \pz \tau_{1}/M$, see Eqs.(\ref{eq:free}), (\ref{eq:meanx}) and (\ref{eq:momav}) for $k=0$.
Moreover, the unitary evolution
introduces an imaginary component in $\sig_t$, according to $\sig_{t_1} = \sig + i \hbar \tau_1 / M$.
Because of such an imaginary component, the jump at time $t_1$ and position $y$ actually modifies the mean
value of the momentum, which after the jump will be:
\begin{equation}
\langle P' \rangle_{t_1} = \pz + \frac{\hbar \, \text{Im}[f_{\sig_{t_1}}]}{\sig^{'\text{R}}_{t_1}} ( \langle X \rangle_{t_1} - y) = 
 \pz + \frac{\hbar M \tau_1}{\hbar^2 \tau^2_1 + M^2 \sig ( \rc^2+\sig)} (y- \langle X \rangle_{t_1}),\label{eq:lpp}
\end{equation}
where $g_{\sig}$ has reduced to $f_{\sig}$, see Eq.(\ref{eq:fsig}), since we are now considering the limit $k=0$, i.e., the original GRW model.
The system varies its momentum proportionally to the distance between the position of the jump and the mean value of the position before the jump. 
This shift of the momentum, in turn, contributes to the change in position after the jump: between the first and
the second jump the mean value of the position evolves as  
$\langle X \rangle_{t} = \langle X' \rangle_{t_1} + \langle P' \rangle_{t_1} (t-t_1)/M$. The iteration of these two transformations,
according to the different spatial distribution of the jumps, will generate some trajectories such that both
the mean position and the mean momentum will diverge to $+\infty$, and some other trajectories where
they will diverge to $-\infty$: in both cases the mean kinetic energy will asymptotically diverge. 
Due to the symmetric probability distribution of the location of each jump,
see for example Eq.(\ref{eq:pygrw}), the effect on the mean momentum now described will be on average null,
and then $\mathbbm{E}[\langle P\rangle_t] = \pz$. However, 
the statistical average of the squared mean value of the momentum will diverge, $\mathbbm{E}[\langle P \rangle_t^2] \rightarrow + \infty$,
and thus the average of the mean energy will diverge with it. 

The introduction of a dissipative mechanism, through a small $k\neq0$, only slightly modifies the action of the jump operators, see Sec. \ref{sec:egm} and Appendix \ref{sec:gs}.
Nevertheless, this tiny modification is enough to damp the long-time momentum fluctuations,
thus leading to an asymptotic finite value of the energy, as will be shown and discussed in Sec.\ref{sec:therm}.

\section{Master equation}\label{sec:me}

Up to now, we have dealt with the stochastic evolution of the wavefunction, as fixed by Eq.(\ref{eq:edsnl}).
The latter provides a complete characterization of the collapse model, as it yields all the possible piecewise deterministic trajectories
which can be obtained according to items 1-4 in Sec. \ref{sec:egm}. Nevertheless, it is often
convenient to study the predictions of the model related with the statistical mean of relevant physical quantities,
i.e., compare with Eq.(\ref{eq:avo}),
\begin{equation}
\av{O}_t \equiv \mathbbm{E}[\langle O \rangle_t]  = \mathbbm{E}[\bra{\psi(t)}\widehat{O}\ket{ \psi(t)}] = \text{Tr}\left\{\mathbbm{E}[\ket{\psi(t)}\bra{\psi(t)}] \widehat{O}\right\} = 
 \text{Tr}\left\{\hat{\rho}(t) \widehat{O}\right\} .\label{eq:avo2}
\end{equation} 
Here, we have introduced 
\begin{equation}\label{eq:erho}
\hat{\rho}(t) \equiv \mathbbm{E}[\ket{\psi(t)}\bra{\psi(t)}], 
\end{equation}
which is by construction a statistical operator on the Hilbert space
associated with the system. Incidentally, since the stochastic wavefunction $\ket{\psi(t)}$
is uniquely determined by the trajectories $\omega_t$, the statistical mean in Eq.(\ref{eq:erho})
corresponds to the mean over the different trajectories $\omega_t$, each one weighted with
its $\mathbbm{P}$-probability density. In the following,
we will focus on the evolution of the statistical operator $\hat{\rho}(t)$,
which will allow us to describe the evolution of relevant physical quantities,
as well as to further characterize the dissipation and the amplification mechanism
in the model.

\subsection{From the stochastic differential equation to the master equation}\label{sec:edsme}
The equation of motion satisfied by $\hat{\rho}(t)$, i.e., the master equation
associated with the extended GRW model, is easily determined by using the product rule
\begin{equation}\label{eq:preme}
\mathd (\ket{\psi(t)} \bra{\psi(t)}) = (\mathd \ket{\psi(t)}) \bra{\psi(t)} + \ket{\psi(t)}(\mathd \bra{\psi(t)}) + (\mathd \ket{\psi(t)})(\mathd \bra{\psi(t)})
\end{equation}
and Eqs.(\ref{eq:edn1})-(\ref{eq:edn3}). Explicitly, Eqs.(\ref{eq:edn1}) and (\ref{eq:edn2}) imply that the stochastic differential equation (\ref{eq:edsnl}) gives
(using the notation $\widehat{L}_y \equiv L(\widehat{X}, \widehat{P})$)
\begin{equation}
\mathd  (\ket{\psi(t)} \bra{\psi(t)}) = -\frac{i}{\hbar} \left[\widehat{H},  \ket{\psi(t)} \bra{\psi(t)} \right] \mathd t + \int \mathd y \left(\frac{\widehat{L}_y 
  (\ket{\psi(t)} \bra{\psi(t)})\widehat{L}^{\dag}_y }{\|\widehat{L}_y  \ket{\psi(t)}\|^2} - \ket{\psi(t)} \bra{\psi(t)}\right) \mathd N_y(t),
\end{equation}
which can be written as \cite{Barchielli1991,Barchielli1994}
\begin{eqnarray}
\mathd  (\ket{\psi(t)} \bra{\psi(t)}) 
&=& -\frac{i}{\hbar} \left[\widehat{H}, \ket{\psi(t)} \bra{\psi(t)} \right] \mathd t+ \lambda \left( \int \mathd y \widehat{L}_y  (\ket{\psi(t)} \bra{\psi(t)}) \widehat{L}^{\dag}_y 
 - \ket{\psi(t)} \bra{\psi(t)} \| \widehat{L}_y  \ket{\psi(t)} \|^2 \right) \mathd t  \nonumber\\
&& +  \int \mathd y \left(\frac{\widehat{L}_y  (\ket{\psi(t)} \bra{\psi(t)}) \widehat{L}^{\dag}_y }{ \| \widehat{L}_y  \ket{\psi(t)} \|^2} - \ket{\psi(t)} \bra{\psi(t)}\right)\left( \mathd N_y(t) - \lambda \| \widehat{L}_y  \ket{\psi(t)} \|^2 \mathd t \right).
\label{eq:prelo}
\end{eqnarray}
Since the defining properties of the conditional expected value imply $\mathbbm{E}[\mathbbm{E}[\ket{\psi(t)} \bra{\psi(t)}| \omega_t]] = \mathbbm{E}[\ket{\psi(t)} \bra{\psi(t)}]$,
we can get an equation for $\hat{\rho}(t)$ by taking the expectation of Eq.(\ref{eq:prelo}) conditioned upon the trajectory $\omega_t$.
The conditional expected value of $\mathd N_y(t)$, see Eq.(\ref{eq:edn3}), implies that the second line in Eq.(\ref{eq:prelo})
does not give any contribution. By exploiting Eq.(\ref{eq:norml}) and further taking the stochastic average,
we end up with the master equation
\begin{eqnarray}\label{eq:megrwdiss}
\frac{\mathd}{\mathd t}\hat{\rho}(t) &=& - \frac{i}{\hbar}\left[\widehat{H} \,,\, \hat{\rho}(t)\right] + \lambda\left(\int \mathd y \, L_y(\widehat{X}, \widehat{P}) \hat{\rho}(t) L^{\dag}_y(\widehat{X}, \widehat{P})  -\hat{\rho}(t) \right) \nonumber\\
&=&   - \frac{i}{\hbar}\left[\widehat{H} \,,\, \hat{\rho}(t)\right] + \lambda\left( \frac{\rc (1+k)}{\sqrt{\pi} \hbar}
\int  \mathd \mom \, e^{\frac{i}{\hbar} \mom \widehat{X}} 
e^{-\frac{\rc^2}{2\hbar^2} \left((1+k) \mom+ 2 k \widehat{P}\right)^2} \hat{\rho}(t) 
e^{-\frac{\rc^2}{2\hbar^2} \left((1+k) \mom+ 2 k \widehat{P}\right)^2}e^{-\frac{i}{\hbar} \mom \widehat{X}}   -\hat{\rho}(t) \right). \nonumber \\
\end{eqnarray}
Indeed, this is a Lindblad master equation \cite{Lindblad1976},
which reduces to the master equation associated with the GRW model
in the limit $k \rightarrow 0$ \cite{Vacchini2007}.

\paragraph*{Role of the normalization condition in Eq.(\ref{eq:norml})}
In the next two paragraphs, we will discuss more in detail the physical meaning of Eq.(\ref{eq:megrwdiss});
before that, let us make the following remark.
Suppose to define a collapse model via an equation as Eq.(\ref{eq:edsnl}), but with
different jump operators $\tilde{L}_y(\widehat{X}, \widehat{P})$,
which do not satisfy
Eq.(\ref{eq:norml}). First, this would imply a total rate of localization dependent on the state of the system
and proportional to
 $\int \mathd y \|\tilde{L}_y(\widehat{X}, \widehat{P})\ket{\psi} \|^2$,
as well as a probability density for the localization position defined as $p(y) =  \|\tilde{L}_y(\widehat{X}, \widehat{P})\ket{\psi} \|^2/(\int \mathd y \|\tilde{L}_y(\widehat{X}, \widehat{P})\ket{\psi} \|^2)$,
compare with Eq.(\ref{eq:pyd}).
Even more importantly, 
as can be easily seen by repeating the calculations of this paragraph,
Eq.(\ref{eq:megrwdiss}) should be replaced with
\begin{eqnarray}
\frac{\mathd}{\mathd t}\hat{\rho}(t) &=& - \frac{i}{\hbar}\left[\widehat{H} \,,\, \hat{\rho}(t)\right] + \lambda\left(\int \mathd y \tilde{L}_y(\widehat{X}, \widehat{P})\hat{\rho}(t)\tilde{L}^{\dag}_y(\widehat{X}, \widehat{P})
 -  \mathbbm{E}\left[ \|\tilde{L}_y(\widehat{X}, \widehat{P})\ket{\psi} \|^2 \ket{\psi(t)}\bra{\psi(t)}\right] \right).
\end{eqnarray}
In order to get a closed equation for the statistical operator $\hat{\rho}(t)$, which would also be in the Lindblad form,
one has to change the stochastic differential equation (\ref{eq:edsnl}) by replacing
the Hamiltonian $\widehat{H}$
with the effective non-hermitian Hamiltonian 
$$\widehat{H}_{\text{eff}} = \widehat{H} - \frac{i}{2} \int \mathd y \tilde{L}_y^{\dag}(\widehat{X}, \widehat{P})\tilde{L}_y(\widehat{X}, \widehat{P}),$$
as well as adding a term which  guarantees the norm preservation of the wavefunction \cite{Barchielli1991}. 
The evolution between the jumps is then no longer unitary,
but it is given by a more general completely positive map.
We conclude that the choice of the jump operators, and, in particular, their property expressed by Eq.(\ref{eq:norml}), is crucial
to define the collapse model in terms of the usual unitary evolution interrupted by sudden jumps, at least as
long as we want the evolution of $\hat{\rho}(t)$ to be described by
a closed linear master equation.

\subsection{Physical meaning of the master equation}
In order to understand the meaning of the master equation (\ref{eq:megrwdiss}),
let us first note that a master equation of the same form appears within the description of the collisional decoherence, and also recall what has been said
at the end of Sec. \ref{sec:amaed}.
Explicitly, the dynamics of a test particle interacting through collisions with a free low density background gas
in the weak coupling regime (and restricting to the one dimensional case for the sake of comparison),
can be characterized through the equation \cite{Vacchini2000,Vacchini2001}:
\begin{equation} 
\frac{\mathd}{\mathd t}\hat{\rho}(t) = - \frac{i}{\hbar}\left[\widehat{H} \,,\, \hat{\rho}(t)\right] + (2 \pi)^2 n_{\text{gas}} 
\int \mathd \mom |\tilde{t}(\mom)|^2 \left( \, e^{\frac{i}{\hbar} \mom \widehat{X}} 
\sqrt{S(\mom, \widehat{P})} \hat{\rho}(t) \sqrt{S(\mom, \widehat{P})} e^{-\frac{i}{\hbar} \mom \widehat{X} } 
 - \frac{1}{2}\left\{S(\mom, \widehat{P})  \, \hat{\rho}(t)\right\} \right). \label{eq:medic}
\end{equation}
Here, $n_{\text{gas}}$ is the density of the gas, while $\tilde{t}(\mom)$ is the Fourier transform of the two-body interaction
potential between the test particle and the gas particles. Finally, $S(\mom, P)$ is a two point correlation function,
which is usually called dynamic structure factor and the operator valued function $S(\mom, \widehat{P})$
is defined through the relation $S(\mom, \widehat{P})\ket{P} = S(\mom, P) \ket{P}$. 
The dynamic structure factor describes
the energy and momentum exchange between the test particle and the gas, and for a free gas of Maxwell-Boltzmann
particles it can be written as
\begin{equation}\label{eq:dsf}
S(\mom, P) = \sqrt{\frac{\beta m}{2 \pi }} \frac{1}{ |\mom|} e^{-\frac{\beta}{8 m}\left(\mom + 2 m E(\mom, P)/ \mom\right)^2},
\end{equation}
where $\beta$ is the inverse temperature and $m$ the mass of the gas particles, while $E(\mom, P) = \frac{\mom^2}{2 M} + \frac{P \mom}{M}$
is the energy exchanged in a collision such that $P \rightarrow P + \mom$. 
Now, if we choose an interaction potential $t(x) = K |x|^{-3/2}$, implying
\begin{equation}
\tilde{t}(\mom) = K \int \frac{\mathd x}{2 \pi \hbar} e^{-\frac{i}{\hbar} \mom x} |x|^{-3/2} =- \frac{K \sqrt{2 |\mom|}}{\sqrt{\pi}\hbar^{3/2}},
\end{equation}
Eq.(\ref{eq:medic}) exactly corresponds to Eq.(\ref{eq:megrwdiss}) upon performing the following identifications
between the parameters of the two equations:
\begin{eqnarray}
\rc &\longleftrightarrow& \frac{\sqrt{2 \pi \beta \hbar^2/m}}{4 \sqrt{\pi} } = \frac{\lambda_{\text{th}}}{4 \sqrt{\pi}} \nonumber\\
\lambda &\longleftrightarrow& \frac{16 \pi K^2 n_{\text{gas}} m}{\hbar^3} \nonumber\\
\newp &\longleftrightarrow&  v_{\text{mp}}. \label{eq:co}
\end{eqnarray}
The new parameter
of the model $\newp$ precisely corresponds to the most probable
velocity of the gas particles  $v_{\text{mp}} = \sqrt{2/(\beta m)}$. In addition, the localization width is fixed by the thermal wavelength $\lambda_{\text{th}}$
of the gas particle. The same correspondence is present between the original GRW model
and the collisional master equation without dissipation \cite{Vacchini2007}. 
Finally, note that the master equation (\ref{eq:medic}), and then Eq.(\ref{eq:megrwdiss}) as well,
fall into the class of translation-covariant Lindblad master equation, whose full characterization was given by Holevo \cite{Holevo1993a,Holevo1993b,Vacchini2009}.

Now that the link with a collisional model has been fixed, the interpretation of Eq.(\ref{eq:megrwdiss})
is quite straightforward.  The Lindblad operators, which describe the
action of the environment on the system, are made up of two terms. The boost operator $\exp(i \mom \widehat{X}/\hbar)$ describes the exchange of a momentum $Q$ 
between the test particle and the background gas, according to $\exp(i \mom \widehat{X}/\hbar) \ket{P} = \ket{P+\mom}$.  
The operator 
\begin{equation}\label{eq:lmw}
\mathbbm{L}(\mom, \widehat{P})\equiv \sqrt{ \frac{\rc (1+k)}{\sqrt{\pi}\hbar}}e^{-\frac{\rc^2}{2\hbar^2}\left( (1+k) \mom+ 2 k \widehat{P} \right)^2}
\end{equation}
provides
the probability amplitude that the change of momentum is equal to $\mom$
if the test particle has momentum $P$, as seen by $ \mathbbm{L}(\mom, P) = \bra{P+Q} \exp(i \mom \widehat{X}/\hbar) \mathbbm{L}(\mom, \widehat{P})\ket{P}$.
By taking $k \rightarrow 0$,
the dependence on the momentum operator
$\widehat{P}$ in Eq.(\ref{eq:lmw}) disappears: the probability density that the system undergoes a momentum variation
$Q$ is fixed, i.e., it does not depend on the momentum of the
system itself. Therefore, the mean value of the momentum is constant, while the momentum variance steadily increases in time. 
Any dissipative effect is excluded from the dynamics, thereby leading to a divergence of the energy \cite{Bassi2005}. 
The master equation (\ref{eq:medic}) reduces to the master equation introduced in \cite{Gallis1990}, which accounts for the recoil-free decoherence dynamics
of a massive particle in the limit $M \rightarrow \infty$
and reproduces the original GRW master equation \cite{Vacchini2007}.
The introduction of
a dependence on $\widehat{P}$ within the Lindblad operators is just what introduces the energy relaxation in the collisional dynamics of
the test particle, thus keeping the mean value of the energy finite, see Sec. \ref{sec:therm}.

As well-known, each master equation in the Lindblad form can be obtained as the statistical mean
of infinite different stochastic differential equations. This fact is often conveyed by saying
that any Lindblad master equation has infinite different unravellings \cite{Carmichael1993,Gardiner1999,Breuer2002,Barchielli2009}, essentially
one for each way of writing it in terms of different Lindblad operators. In particular, the master equation of our model
could be obtained also by starting from a stochastic differential equation
with a family of counting processes $\left\{N_{\mom}\right\}_{\mom\in \mathbbm{R}}$,
one for each variation of the system's momentum, and with jump operators
given by $\exp(i \mom \widehat{X}/\hbar) \mathbbm{L}(\mom, \widehat{P})$, i.e. by the Lindblad operators in Eq.(\ref{eq:megrwdiss}).
This kind of unravelling was introduced in \cite{Breuer2007,Busse2010} to study numerically the solutions of the master equation.
However, one should keep in mind that the stochastic differential equation
can be associated with a collapse model only if the localization mechanism is present. 
The position spread of the wavefunction has to be reduced in 
the different trajectories of the model. In a nutshell, not every unravelling of the master equation can be a candidate
to describe a collapse model. One can easily see that jump
operators such as $\exp(i \mom \widehat{X}/\hbar) \mathbbm{L}(\mom, \widehat{P})$
would not induce a localization of the wavefunction, while their Fourier transform
defines the jump operators of our model, i.e.,
\begin{equation}
L_y(\widehat{X}, \widehat{P}) = \int \frac{\mathd \mom}{\sqrt{2 \pi \hbar}}e^{-\frac{i}{\hbar} \mom y} e^{\frac{i}{\hbar} \mom \widehat{X}} \mathbbm{L}(\mom, \widehat{P}).
\end{equation}

\subsection{Solution in the position representation}{\label{sec:sitpe}}

The solution of the master equation can be obtained by exploiting the characteristic function \cite{Savage1985,Smirne2010}
\begin{equation}\label{eq:chid}
\chi (\nu,\mu,t)=\text{Tr}\left\{\hat{\rho}(t)
\,e^{\frac{i}{\hbar}\left(\nu \hat{X}+\mu \hat{P}\right)}\right\},
\end{equation}
as the matrix elements of the statistical operator in the position representation
can be obtained through
\begin{equation} \label{eq:inv}
 \rho(X,X',t) = \int \frac{\mathd \nu}{2 \pi \hbar} e^{- i \nu (X+X')/(2 \hbar)} \chi(\nu, X-X',t). 
\end{equation}
In appendix \ref{app:dote}, we show that Eq.(\ref{eq:megrwdiss}) is equivalent to the following
equation for the characteristic function:
\begin{equation}\label{eq:char}
\partial_t \chi (\nu,\mu,t) = \frac{\nu}{M} \partial_{\mu} \chi (\nu,\mu,t) + \lambda \chi \left(\nu,\mu\left(\frac{1-k}{1+k}\right),t\right)\exp\left(-\frac{\nu^2 \rc^2 k^2}{\hbar^2}-\frac{ \mu^2}{4 \rc^2(1+k)^2}\right)
-\lambda \chi (\nu,\mu,t).
\end{equation}
Since $k = \hbar/(2M \newp \rc) \ll 1$, see Eq.(\ref{eq:kg}), in general the off-diagonal elements $\rho(X,X',t)$ will not vary significantly by replacing $X-X'$
with $(X-X')(1-k)/(1+k)$, while keeping $X+X'$ constant.
Under this condition, one can neglect the dependence on $k$ within the second term at the right hand side of Eq.(\ref{eq:char}), see Eq.(\ref{eq:inv}),
thus getting
\begin{equation}\label{eq:char2}
\partial_t \chi (\nu,\mu,t) = \frac{\nu}{M} \partial_{\mu} \chi (\nu,\mu,t) + \lambda\left( \Phi(\nu,\mu) -1\right) \chi (\nu,\mu,t),
\end{equation}
with
\begin{equation}
\Phi(\nu,\mu) \equiv \exp\left(-\frac{\nu^2 \rc^2 k^2}{\hbar^2}-\frac{ \mu^2}{4 \rc^2(1+k)^2}\right).
\end{equation}
This first order partial differential equation is solved by \cite{Smirne2010}
\begin{eqnarray}
  \chi(\nu, \mu, t) & = & \chi^0(\nu, \nu t / M +\mu, t) e^{-  \lambda \int^t_0 (1 - \Phi(\nu, \nu(t - t') / M
  +\mu)) \mathd t'},  \label{eq:charsol}
\end{eqnarray}
where the function $\chi^0(\nu, \nu t / M +\mu, t) $ satisfies the free equation $\partial_t \chi^0(\nu, \nu t / M +\mu, t) = 
(\nu/M)  \partial_{\mu}  \chi^0(\nu, \mu, t)$.
By using Eq.(\ref{eq:inv}),
one ends up with the solution of the master equation in the position representation,
\begin{equation}
  \rho (X, X', t) = \int
  \frac{\mathd s \mathd \nu}{(2 \pi \hbar)} e^{-
  i\nu s/ \hbar} e^{-\lambda \int^t_0 (1
  - \Phi (\nu, \nu(t - t') / M +X-X'))\mathd t'} \rho^0 (X+s, X' +s, t),\label{eq:mesol}
\end{equation}
where $\rho^0 (X, X', t)$ is the solution of the free Schr{\"o}dinger equation.
Note that $\Phi(0,0) = 1$ guarantees the trace preservation of $\rho(t)$, while $\Phi(\nu,\mu) = \Phi(-\nu,-\mu)$
guarantees its hermiticity.

By means of the expression of the statistical operator at time $t$,
we can explicitly evaluate the dynamics of relevant physical quantities, such as
the statistical mean value of the position, see Eq.(\ref{eq:avo2}),
as well as its variance. 
The variance $(\Delta_{\rho_t} O)^2$ of the observable represented by $\widehat{O}$, 
if the average state of the system is $\hat{\rho}(t)$, reads
\begin{equation}\label{eq:frak}
(\Delta_{\rho_t} O)^2 \equiv \text{Tr}\left\{\widehat{O}^2\hat{\rho}(t)\right\}-\text{Tr}^2\left\{\widehat{O}\hat{\rho}(t)\right\}.
\end{equation}
For the position mean value, by Eq.(\ref{eq:mesol}) we have
\begin{eqnarray}
\av{X}_t  &= &\int \frac{\mathd s \mathd \nu \mathd X}{(2 \pi \hbar)} (X-s) e^{-
  i\nu s/ \hbar} e^{-\lambda \int^t_0 (1
  - \Phi (\nu, \nu(t - t') / M ))\mathd t'} \rho^0 (X, X, t) \nonumber\\
  &=& \av{X}_t^S+ i \hbar \partial_{\nu}e^{-\lambda \int^t_0 (1
  - \Phi (\nu, \nu(t - t') / M ))\mathd t'} |_{\nu=0} = \av{X}_t^S,
\end{eqnarray}
where $\av{X}_t^S$ denotes the mean value under free Schr{\"o}dinger evolution.
As for the case without dissipation \cite{Ghirardi1986}, the average effect of the localization processes
does not influence the evolution of the mean value of the position.
Analogously, one finds
\begin{eqnarray}
\av{X^2}_t &=& \int \frac{\mathd s \mathd \nu \mathd X}{(2 \pi \hbar)} (X-s)^2 e^{-
  i\nu s/ \hbar} e^{-\lambda \int^t_0 (1
  - \Phi (\nu, \nu(t - t') / M ))\mathd t'} \rho^0 (X, X, t) \nonumber\\
  &=& \av{X^2}_t^S+ 2 i \hbar \partial_{\nu}e^{-\lambda \int^t_0 (1
  - \Phi (\nu, \nu(t - t') / M ))\mathd t'} |_{\nu=0}\, \av{X}_t^S -\hbar^2 \partial^2_{\nu}e^{-\lambda \int^t_0 (1
  - \Phi (\nu, \nu(t - t') / M ))\mathd t'} |_{\nu=0} \nonumber\\
  &=& \av{X^2}_t^S +2 k^2 \rc^2 \lambda t + \frac{\hbar^2 \lambda}{6 \rc^2(1+k)^2 M^2} t^3.
\end{eqnarray}
Thus, the position variance is, see Eq.(\ref{eq:frak}),
\begin{equation}
(\Delta_{\rho_t} X)^2  =(\Delta^S_{\rho_t} X)^2  +2 k^2 \rc^2 \lambda t + \frac{\hbar^2 \lambda}{6 \rc^2(1+k)^2 M^2} t^3,
\end{equation}
where $(\Delta^S_{\rho_t} X)^2$ is the position variance under the Schr{\"o}dinger evolution.
The dissipation introduces a spread of the position variance which is linear in time,
in addition to the term proportional to $t^3$, already present in the original GRW model \cite{Ghirardi1986}.
It is worth noting that the position variance referred to the statistical operator $\hat{\rho}(t)$ is not the
statistical average of the position variance for the stochastic wavefunction, see Eq.(\ref{eq:pv}).
While for the mean values of the observables one has Eq.(\ref{eq:avo2}), an analogous relation
does not hold for the variances, so that in general 
$
\mathbbm{E}[(\Delta_{\psi_t} O)^2]  \neq (\Delta_{\rho_t} O)^2
$. The reason for that is essentially that the stochastic average $\mathbbm{E}$ and the square of the trace do not commute \cite{note2}, see Eq.(\ref{eq:frak}).

\section{Time evolution of the mean energy}\label{sec:therm}

By virtue of the dynamics for the statistical operator,
we can now show in an explicit way how the modification
of the jump operators leads to energy relaxation. To do so, we
study the evolution in time of the mean energy $\av{H}_t$,
focussing on the case $\widehat{H} = \widehat{P}^2/(2M)$.

Instead of using the solution in the position representation given by Eq.(\ref{eq:mesol}),
we can compute directly the dynamics of the mean value of any operator $f(\widehat{P})$,
which is a function of the momentum operator only, as follows.
Since $\left[\widehat{H}, f(\widehat{P})\right] = 0$ and $\exp(-i \mom \widehat{X}/\hbar) f(\widehat{P})\exp(i \mom \widehat{X}/\hbar) = f(\widehat{P} + \mom)$,
Eq.(\ref{eq:megrwdiss}) implies
\begin{eqnarray}
\frac{\mathd}{\mathd t} \av{f(P)}_t&=&\lambda \left(  \frac{\rc(1+k)}{\sqrt{\pi}\hbar} \int \mathd \mom\,
\av{e^{-\rc^2 \left((1+k)\mom+2 k P\right)^2/\hbar^2} f(P+\mom)}_t-\av{f(P)}_t \right) \nonumber\\
&=&\lambda   \frac{\rc(1+k)}{\sqrt{\pi}\hbar} \int \mathd \mom\,
\av{e^{-\rc^2 \left((1+k)\mom+2 k P\right)^2/\hbar^2}\left(f(P+\mom)-f(P)\right)}_t.
\end{eqnarray}
It is worth mentioning how this result is a direct consequence of the translation covariance of
the master equation \cite{Vacchini2009}.
For the momentum operator, one has 
\begin{eqnarray}
\frac{\mathd}{\mathd t} \av{P}_t&=&\lambda   \frac{\rc(1+k)}{\sqrt{\pi}\hbar} \int \mathd \mom\,
\mom \av{e^{-\rc^2 \left((1+k)\mom+2 k P\right)^2/\hbar^2}}_t = - \frac{2k}{k+1} \lambda \av{P}_t.
\end{eqnarray}
The mean value of the momentum is then damped exponentially in time with the rate given by the product between the GRW rate $\lambda$
and the dissipation parameter $2k/(k+1)$; namely
\begin{equation}\label{eq:pt}
\av{P}_t = e^{- \frac{2k \lambda}{k+1} t} \av{P}_0.
\end{equation}
By removing dissipation, i.e. setting $k=0$, one recovers a constant value of the mean momentum.
For the kinetic energy, one has
\begin{eqnarray}
\frac{\mathd}{\mathd t} \av{H}_t&=&\lambda   \frac{\rc(1+k)}{2 \sqrt{\pi}\hbar M} \int \mathd \mom\,
\av{e^{-\rc^2 \left((1+k)\mom+2 k P\right)^2/\hbar^2}\left(\mom^2 + 2 P Q \right)}_t \nonumber\\
&=& \frac{\hbar^2 \lambda}{4 M \rc^2(1+k)^2} - \frac{4\lambda k}{(1+k)^2} \av{H}_t,
\end{eqnarray}
so that
\begin{equation}\label{eq:ht}
\av{H}_t = \left(\av{H}_0 - H_{\text{as}}\right) e^{-\xi t} + H_{\text{as}}:
\end{equation}
the mean value of the energy relaxes with rate
\begin{equation}\label{eq:rrate}
\xi =  \frac{4\lambda k}{(1+k)^2}
\end{equation}
to the asymptotic value
\begin{equation}\label{eq:has}
H_{\text{as}} = \frac{\hbar^2}{16 M \rc^2 k}.
\end{equation}
Indeed, since $\av{H}_t = \mathbbm{E}[\langle H \rangle_t]$ reaches an asymptotic finite
value, $\langle H \rangle_t$ for $t\rightarrow \infty$ will be almost surely finite on the trajectories of the collapse model, see also Sec. \ref{sec:lmaei}.

The asymptotic value of the mean energy given by Eq.(\ref{eq:has}) corresponds to a temperature of the noise \cite{Bassi2005}
\begin{equation}
T = \frac{\hbar^2}{8 k_B M \rc^2 k} =  \frac{\hbar \newp}{4 k_B \rc} \approx 10^{-1} K,\label{eq:T}
\end{equation}
where we exploited the definition of $k$ in Eq.(\ref{eq:kg}).  Most importantly,
we have obtained a value of the noise temperature which is independent from the mass
of the system. Let us stress that this is a consequence of the choice of the
jump operator $L_y(\widehat{X}, \widehat{P})$ in Eq.(\ref{eq:lyxp2}), including 
the specific dependence on the new parameter $\newp$.
In addition, the estimate in Eq.~(\ref{eq:T}) justifies our initial choice for the numerical value of $\newp$: 
it yields the same order of magnitude of the
temperature as the continuous collapse model analyzed in \cite{Bassi2005}. 
Crucially, this value also points out how 
a proper collapse noise does not need suspicious and ad hoc properties.
A classical noise with typical cosmological features (low temperature)
can guarantee the collapse of the wavefunction,
along with the thermalization to a finite energy \cite{Bassi2010b}.
Finally, as expected, in the limit $\newp \rightarrow \infty$
we recover an infinite temperature of the noise, corresponding to the original GRW model.

The dissipation rate is $\xi$, see Eqs. (\ref{eq:ht}) and (\ref{eq:rrate})
and therefore its ratio with the collapse rate is  
\begin{equation}
\frac{\xi}{\lambda} =\frac{4k}{(1+k)^2} \ll 1,
\end{equation}
The condition $k\ll1$ guarantees that the collapse occurs on a time scale $\lambda^{-1}$
much shorter than the time scale $\xi^{-1}$ of dissipation.
In addition, recall that the collapse rate $\lambda$ is not modified
by the introduction of the dissipation, see Secs. \ref{sec:egm} and \ref{sec:this}.
Any experimental investigation on the effects of collapse models on short time scales
will thus not be able to highlight a significant role of the extended dissipative GRW model with respect
to the original one. On the other hand, the exponential relaxation
of the energy to a finite value drastically changes
the predictions of the model
for the experiments
which involve the secular behavior of the energy \cite{Adler2007}.
The most important example is provided by the heating of the intergalactic medium (IGM),
which up to now yields the second strongest upper bound to the localization rate \cite{Adler2007,Adler2009}.
Such a bound has been derived by considering the continuous spontaneous localization (CSL) collapse model,
which allows to deal with the Fermi or Bose statistics for identical particles.
Hence, the influence of dissipation on the predictions on the secular behavior of
the energy has been evaluated in \cite{Smirne2014}, where we extended the CSL model
in order to include dissipation.

It is worth noting how the asymptotic value of the energy is fixed by the threshold value introduced in Sec. \ref{sec:egm}, see Eq.(\ref{eq:soglia}).
Recall that the jump operators cease to induce a localization of the wavefunction for gaussian wave packets with a position spread smaller than
\begin{equation}\label{eq:thrd}
(\Delta_{\phi} X)^2_{\text{thr}} \equiv 2 \rc^2 k.
\end{equation}
As already noticed, the Hamiltonian part of the master equation
does not give any contribution to
the evolution equation of the mean value of the energy. This means that the same equation would be obtained by starting
from a master equation as Eq.(\ref{eq:megrwdiss}), but without the Hamiltonian term \cite{nota3}.
Such a master equation can be thought as given by the statistical mean of the stochastic differential equation, compare
with Eq.(\ref{eq:edsnl}),
\begin{equation}\label{eq:aux}
\mathd \ket{\psi(t)} =\int \mathd y \left(\frac{L_y(\widehat{X}, \widehat{P})}{\|L_y(\widehat{X}, \widehat{P}) \ket{\psi(t)}\|} - \mathbbm{1}\right) \ket{\psi(t)} \mathd N_y(t).
\end{equation}
The trajectories provided by this equation
are fixed by the action of the jumps operators only. Now, for the sake of simplicity, consider
an initial minimum uncertainty gaussian state as in Eq.(\ref{eq:gaus}). Its position variance
$(\Delta_{\phi_0} X)^2 = \sig /2$ is increasingly contracted by the action of the jumps operator, see Eq.(\ref{eq:xps}), until it reaches the threshold
value in Eq.(\ref{eq:thrd}), which represents the asymptotic value of  $(\Delta_{\phi_t} X)^2$ for the trajectories of Eq.(\ref{eq:aux}).
But then, since the initial minimum uncertainty gaussian state 
has remained a minimum uncertainty gaussian state due to the absence of the free Schr{\"o}dinger evolution,
the asymptotic value of the momentum variance will be $(\Delta_{\phi} P)^2_{\text{thr}} = \hbar^2/(4 (\Delta_{\phi} X)^2_{\text{thr}})$.
As the mean value of the momentum $\langle P \rangle_t$ relaxes to zero, $(\Delta_{\phi} P)^2_{\text{thr}}$ determines
the asymptotic value of the mean energy as $H_{\text{as}} = (\Delta_{\phi} P)^2_{\text{thr}}/(2M)$,
which is the value in Eq.(\ref{eq:has}). Let us emphasize once more that this description
does not correspond to what actually happens in the collapse model fixed by Eq.(\ref{eq:edsnl}),
but characterizes an auxiliary model, fixed by Eq.(\ref{eq:aux}), which gives the same
predictions as the extended GRW model, as long as the statistical mean value of the energy is concerned.

\section{Amplification mechanism}\label{sec:ampl}

As recalled in Sec. \ref{sec:amaed}, the amplification mechanism is  a basic  feature of any collapse model.
The localization of the wavefunction due to the jumps has to increase with 
the number of the constituents of the system, as well as with its overall mass. 
In this section, we show that such a mechanism can be proved also in the presence of dissipation,
if a rigid body is taken into account. We also argue that the description
of more complex systems, in which the internal dynamics plays a significant role, calls
for a more realistic and detailed characterization of the $N$-particle evolution than that
regularly used in collapse models.

\subsection{Master equation for the center of mass of a rigid body}\label{sec:mef}

Let us start by dealing with an $N$-particle system subject to localization processes occurring
individually for each constituent, which
means that the overall effect of the noise consists simply in the sum
of the independent effects on the single particles, see Sec. \ref{sec:amaed}. In the next paragraph, we will see how the validity of
this assumption breaks down if the internal motion
of the system has to be taken into account.
Hence, we consider now the master equation for the total state $\hat{\varrho}(t)$
given by, compare with Eq.(\ref{eq:megrwn}) [we are using the analogous short-hand notation],
\begin{equation}\label{eq:megrwndiss}
\frac{\mathd}{\mathd t}\hat{\varrho}(t) = - \frac{i}{\hbar}\left[\widehat{H}_{\text{T}} \,,\, \hat{\varrho}(t)\right] 
+ \sum_j \lambda_j \left(\int \mathd y \, L_y(\widehat{X}_j, \widehat{P}_j) \hat{\varrho}(t) L^{\dag}_y(\widehat{X}_j, \widehat{P}_j)  -\hat{\varrho}(t) \right),
\end{equation}
where $\widehat{X}_j$ ($\widehat{P}_j$) is the position (momentum) operator of the $j$-th particle, see Eq.(\ref{eq:lyxp2}).
Again, we introduce the center of mass coordinates through Eq.(\ref{eq:rj}),
and we assume a total Hamiltonian as in Eq.(\ref{eq:hhh}).
Hence, the state of the center of mass, 
$
\hat{\rho}_{\text{CM}}(t) = \text{Tr}_{\text{REL}}\left\{\hat{\varrho}(t)\right\}
$,
satisfies
\begin{eqnarray}
&&\frac{\mathd}{\mathd t}\hat{\rho}_{\text{CM}}(t) = - \frac{i}{\hbar}\left[\widehat{H}_{\text{CM}} \,,\, \hat{\rho}_{\text{CM}}(t)\right] + 
\sum_j \lambda_j 
\left( \frac{\rc (1+k_j)}{\sqrt{\pi} \hbar}
\int  \mathd \mom \, e^{\frac{i}{\hbar} \mom \widehat{X}_{\text{CM}}}\right. \nonumber\\
&&\left.Tr_{\text{REL}} 
\left\{e^{-\frac{\rc^2}{2\hbar^2} \left((1+k_j) \mom+ 2 k_j \widehat{P}_j\right)^2} \hat{\varrho}(t) 
e^{-\frac{\rc^2}{2\hbar^2} \left((1+k_j) \mom+ 2 k_j \widehat{P}_j\right)^2} \right\}e^{-\frac{i}{\hbar} \mom \widehat{X}_{\text{CM}}}   -\hat{\rho}_{\text{CM}}(t) \right),\nonumber
\end{eqnarray}
where we used the cyclicity of the partial trace over the relative degrees of freedom.
The parameters $k_j$ are proportional to $1/M_j$ according to the relation, see Eq.(\ref{eq:kg}),
\begin{equation}
k_j = \frac{\hbar}{2 M_j \newp  \rc}
\end{equation}
and $k_j \ll 1$. As a consequence, the previous equation can be well approximated by
\begin{eqnarray}
&&\frac{\mathd}{\mathd t}\hat{\rho}_{\text{CM}}(t) = - \frac{i}{\hbar}\left[\widehat{H}_{\text{CM}} \,,\, \hat{\rho}_{\text{CM}}(t)\right] + 
\sum_j \lambda_j 
\left( \frac{\rc}{\sqrt{\pi} \hbar}
\int \mathd \mom \, e^{\frac{i}{\hbar} \mom \widehat{X}_{\text{CM}}}\right. \nonumber\\
&&\left.Tr_{\text{REL}} 
\left\{e^{-\frac{\rc^2}{2\hbar^2} \left( \mom+ 2 k_j \widehat{P}_j\right)^2} \hat{\varrho}(t) 
e^{-\frac{\rc^2}{2\hbar^2} \left(\mom+ 2 k_j \widehat{P}_j\right)^2} \right\}e^{-\frac{i}{\hbar} \mom \widehat{X}_{\text{CM}}}   -\hat{\rho}_{\text{CM}}(t) \right).\nonumber
\end{eqnarray}
Now, we introduce the crucial assumption that we are dealing with a rigid body,
so that $\widehat{P}_j \approx M_j \widehat{P}_{\text T} / M_{\text T}$, where $\widehat{P}_{\text T} = \sum_j \widehat{P}_j$ is the total momentum.
Note that the same assumption plays a relevant role also in order to define a proper measure for the macroscopicity of quantum superpositions \cite{Nimmrichter2013}.
Thus, we finally get
\begin{eqnarray}
\frac{\mathd}{\mathd t}\hat{\rho}_{\text{CM}}(t) &=& - \frac{i}{\hbar}\left[\widehat{H}_{\text{CM}} \,,\, \hat{\rho}_{\text{CM}}(t)\right]  \nonumber\\
&&+\lambda_{\text T} 
\left( \frac{\rc}{\sqrt{\pi} \hbar}
\int  \mathd \mom \, e^{\frac{i}{\hbar} \mom \widehat{X}_{\text{CM}}}
e^{-\frac{\rc^2}{2\hbar^2} \left( \mom+ 2 k_{\text T} \widehat{P}_{\text T}\right)^2} \hat{\rho}_{\text{CM}} (t) 
e^{-\frac{\rc^2}{2\hbar^2} \left(\mom+ 2 k_{\text T} \widehat{P}_{\text T}\right)^2} e^{-\frac{i}{\hbar} \mom \widehat{X}_{\text{CM}}}   -\hat{\rho}_{\text{CM}}(t) \right),\label{eq:ampldiss}
\end{eqnarray}
where, indeed, 
\begin{eqnarray}
\lambda_{\text{T}} &=& \sum_j \lambda_j, \nonumber \\
k_{\text{T}} &=& \frac{\hbar} {2 M_{\text{T}} \newp \rc}.
\end{eqnarray}
This equation is equivalent to the one-particle equation,
see Eq.(\ref{eq:megrwdiss}), with the replacements 
\begin{eqnarray}
\lambda &\rightarrow& \lambda_{\text{T}} \nonumber\\
M &\rightarrow& M_{\text{T}}, \label{eq:repl}
\end{eqnarray} 
apart from the factor $1+k_{\text T}$ which should multiply  $\mom$ in the argument of the square exponential, as well as the coefficient $\rc/(\sqrt{\pi}\hbar)$.
This difference is safely negligible: one has, e.g., $k_{\text{T}} = 5 \times 10^{-29} \ll 1$ for a macroscopic
body of mass $10^{-3} \text{Kg}$. Of course, the remaining dependence
on $k_{\text{T}}$ in Eq.(\ref{eq:ampldiss}) cannot be neglected, since this would
modify the asymptotic behavior of the momentum and energy of the system, see Sec. \ref{sec:therm}.
The center of mass of a rigid body behaves as a single particle of mass $M_{\text{T}}$, but with a rate of localization
increased according to the number of components of the body.
As shown in Sec. \ref{sec:paml}, see especially Fig. \ref{fig:loc},
the replacements in Eq.(\ref{eq:repl}) induce the expected amplification mechanism:
the localization processes do not affect the microscopic systems
on observable time scales, while they allow to treat the center of mass
of a macroscopic rigid body as a classical, well-localized, object.

\subsection{Relevance of the internal dynamics}

The analysis of the previous paragraph can be applied only if the influence
of the internal dynamics on the evolution of the center of mass is negligible.
Actually, we think that the simple procedure usually exploited in order
to show the amplification mechanism,
which describes the evolution of the center of mass through the same master equation
as for the one-particle system with the
replacements in Eq. (\ref{eq:repl}),
has a proper physical meaning just if one takes into account a rigid body. 
In different situations, one would generally need and also expect
a more complex analysis, 
which encompasses the interrelationship between the mutual
interaction of the $N$ particles and the action of the noise. 
We will develop such an analysis in a future and dedicated work.
Once again, the purpose of our investigation should be clear: in our view collapse models are to be understood as phenomenological models
and the limits of the first simpler models can be highlighted and overcome by
searching for more realistic characterizations.  

First, let us examine a simple example, which starts from the study put forward in \cite{Benatti1988}; we will draw, however, different conclusions. 
Consider the two-particle state given by $\ket{\psi} = \ket{\phi^{0,0,\sig}_{\text{CM}}, \phi^{\xz,0,\sig'}_{\text{REL}}}$,
so that
\begin{equation}
\scalar{X_{\text{CM}}, X_{\text{REL}}}{\psi} = C e^{-X_{\text{CM}}^2/(2 \sig)} e^{-(X_{\text{REL}}-\xz)^2/(2 \sig')},  \label{eq:bgrw1}
\end{equation}
with $C = (\pi^2 \sig \sig')^{-1/4}$ the normalization constant.
This is a product state with respect to the partition of the total Hilbert space in terms of the center of mass
and relative degrees of freedom, while it describes an entangled state of the
two particles.
We set $M_1 = M_2$, so that $X_{\text{CM}} = (X_1 + X_2)/2$, while, as usual, 
$X_{\text{REL}} = X_1 - X_2$. Now, suppose that a localization process centered at a position $y$
affects particle 1 and that we can describe such a process through the operator $L_y(\widehat{X}_1, \widehat{P}_1)\otimes \mathbbm{1}_2$.
The localization process is assumed to influence individually each constituent of the $N$-particle
system:  its action on particle 1 is independent from particle 2.
Hence, by using Eq.(\ref{eq:braxL})
one has 
 \begin{eqnarray}
&&\bra{X_{\text{CM}}, X_{\text{REL}}}(L_y(\widehat{X}_1, \widehat{P}_1)\otimes \mathbbm{1}_2) \ket{X'_{\text{CM}}, X'_{\text{REL}}} =
 (\pi  \rc^2(1+k)^2)^{-1/4} 
 \delta\left(X'_{\text{CM}} - \frac{X'_{\text{REL}}}{2}-X_{\text{CM}} + \frac{X_{\text{REL}}}{2} \right)
 \nonumber \\
&& \times e^{-\frac{(X_{\text{CM}}+X_{\text{REL}}/2-y)^2}{2  \rc^2(1+k)^2}}
  \delta\left(\left(\frac{1-k}{1+k}\right)(X_{\text{CM}}+\frac{X_{\text{REL}}}{2})-X'_{\text{CM}}-\frac{X'_{\text{REL}}}{2} +\left(\frac{2k}{1+k}\right) y \right), 
\end{eqnarray}
and therefore the state after the collision, $\ket{\psi_y} = (L_y(\widehat{X}_1, \widehat{P}_1)\otimes \mathbbm{1}_2) \ket{\psi}/\|(L_y(\widehat{X}_1, \widehat{P}_1)\otimes \mathbbm{1}_2) \ket{\psi}\|$, can be written as
\begin{equation}
\scalar{X_{\text{CM}}, X_{\text{REL}}}{\psi_y} = C_y e^{-\frac{(X_{\text{CM}}+ X_{\text{REL}}/2 -y)^2}{2 \rc^2 (1+k)^2}} e^{-\left(\frac{X_{\text{CM}}}{1+k}-\frac{k X_{\text{REL}}}{2(1+k)}+\frac{k y}{1+k} \right)^2/(2 \sig)}
e^{-\left(\frac{X_{\text{REL}}}{1+k}-\frac{2 k X_{\text{CM}}}{1+k}+\frac{2 k y}{1+k}-\xz \right)^2/(2 \sig')},
\end{equation}
which now also includes entanglement between the center of mass and the relative degrees of freedom.
The mean values of the center of mass and relative positions after the localization process are 
\begin{eqnarray}
\langle X_{\text{CM}} \rangle & = & \frac{2(\xz - 2y)((1-k) k \rc^2-\sig)}{4(1-k)^2 \rc^2+4\sig+\sig'} \nonumber \\
\langle X_{\text{REL}}\rangle & = & \frac{4 \xz((1-k)\rc^2+\sig)-2 y(4(1-k)k \rc^2-\sig')}{4(1-k)^2 \rc^2 + 4\sig +\sig'}.\label{eq:meanbgrw}
\end{eqnarray}
Now, we can associate the initial relative state $\ket{\phi^{\xz,0,\sig'}_{\text{REL}}}$
with the ground state of a one-dimensional harmonic oscillator.
In the spirit of \cite{Benatti1988}, 
we can describe 
a two-atom or a two-nucleon ground state by setting properly the wavefunction width $\sig'/2$ .
We take $\sig' = 10^{-22}m^2$ for the atoms and $\sig' = 5 \times 10^{-29}m^2$ for the nucleons,
so that the level spacing $\delta E = \hbar^2/(M \sig')$ of the corresponding harmonic oscillator
is of the order of, respectively, $eV$ and $MeV$, as it must be.
In both cases one has $k \rc^2 \gg \sig'$ and thus, if we also assume $k \rc^2 \gg \sig$,
the mean values in Eq.(\ref{eq:meanbgrw}) can be well approximated as
\begin{eqnarray}
\langle X_{\text{CM}} \rangle & \approx & \frac{k(\alpha-2y)}{2-2k} \nonumber \\
\langle X_{\text{REL}}\rangle & \approx & \frac{\alpha-2 k y}{1-k}. \label{eq:xcmrel}
\end{eqnarray}
Analogously, one can show that the two position variances are left almost unchanged; namely, after the localization
one has
$(\Delta_{\psi_y} X_{\text{CM}})^2\approx \sig/2 + k^2 \sig'/8 $ and $(\Delta_{\psi_y} X_{\text{REL}})^2\approx \sig'/2 + 2 k^2 \sig$.
The probability for the localization process to take place at the position $y$ is
\begin{equation}
p(y) =\| (L_y(\widehat{X}_1, \widehat{P}_1)\otimes \mathbbm{1}_2) \ket{\psi}\|^2 \approx \frac{1}{\sqrt{\pi} \rc (1-k)}e^{-\frac{(y-\xz/2)^2}{\rc^2(1-k)^2}}.
\end{equation}
This provides a gaussian distribution of the localization position, centered around the initial mean position $\xz/2$ of particle 1.
The width  $\rc (1-k)$ of this gaussian distribution, which is independent from the initial state of the system,
implies that there is a non-negligible probability to have a localization process within, say, $10^{-7} m$
away from the initial mean position. But now, let us focus on the case of a two-nucleon state.
According to Eq.(\ref{eq:xcmrel}),
the mean value of the relative position can thus be increased by 
a localization process from $\alpha \approx 10^{-15} m$ up to approximately $10^{-11}m$,
which clearly indicates that $\ket{\psi_y}$ can be no longer
associated with a nucleon bound state. In the original GRW model,
i.e., for $k=0$, in this regime $\langle X_{\text{REL}}\rangle$
would not be modified by the localization process, see Eq.(\ref{eq:xcmrel}).

One could think that the inclusion of dissipation within the GRW model necessarily leads to sudden internal transitions or even dissociations
of nuclei, which would be of course an unacceptable feature of the model.
This was actually the conclusion drawn in \cite{Benatti1988}, where a similar example was considered \cite{nota4}.
We think, however, that this is not the case and that the previous example is pointing out something different.
One should in fact realize that the description of the localization mechanism has been carried out independently from
the presence of a mutual interaction between the components of matter. When one says that
the localization on a 2-particle system is described by the jump operator $L_y(\widehat{X}_1, \widehat{P}_1)\otimes \mathbbm{1}_2$
(and by $\mathbbm{1}_1\otimes L_y(\widehat{X}_2, \widehat{P}_2)$), this means that the
localization process on each particle is the same whether or not they mutually interact. However, it is clear that such a characterization
is in general not realistic and thus possibly leads to unphysical predictions.
In the previous example, one concludes that the two nucleons are shifted apart much farther than the nuclei length scale by the localization mechanism,
simply because the latter has been described without taking into account the effects of the interaction between the two nucleons, e.g. their binding energy.
In order to estimate the average variation of the energy $\Delta E$ due to the localization process, let us consider for simplicity
the action of the localization operator on a one-particle gaussian state $\ket{\phi^{\xz,0,\sig}}$, see  Eq.(\ref{eq:xps}).
Hence, under the assumption $r^2_c \gg \gamma$, the average exchanged energy $\Delta E$ 
is of the order of 
\begin{equation}
\Delta E \approx \frac{\hbar^2 k}{M \sig}. 
\end{equation}
This value is 5 orders of magnitude smaller than the energy needed for an internal transition in the case of a nucleon bound state,
so that it should be clear how the localization processes of the model cannot induce matter dissociation.

Once again, the analogy with decoherence turns out to be useful. If one considers the scattering of a tracer particle
with a particle of its environment, the effects of the collision will be radically different whether or not the tracer particle is in a bound state. 
For instance, think about a gas of free tracer particles interacting with a background gas, in the low density regime: 
to a good extent, the collisions of each tracer particle with the environmental gas can be treated independently,
so that the overall dynamics can be acquired by summing up the individual ones, see Eq. (\ref{eq:megrwndiss}). 
On the other hand, if the tracer particles are in a bound state,
their interaction Hamiltonian will combine with the interaction Hamiltonian between each of them and
the environmental particles, thus modifying the scattering processes. 
More in general, consider the derivation of a master equation for an $N$-particle
open system starting from the $1$-particle master equation, but including a mutual interaction between the $N$ particles during the interaction with the environment.
Usually, the proper master equation cannot be obtained by simply adding the mutual interaction Hamiltonian to the Hamiltonian term and summing up the $1$-particle
Lindblad equations,
but the Lindblad operators should be modified, as well. Actually, in general one cannot even expect to obtain a Lindblad master equation
for the single particles, not to mention the center of mass of the $N$-particle system. A non-Markovian description of these dynamics
will likely come into play  \cite{Smirne2010}. For the interested readers, in \cite{Santos2014} a microscopic derivation of the master equation of 
two interacting qubits in a common environment is given.
This work highlights in a clear way how the mutual interaction between the two qubits modifies significantly the resulting Lindblad structure,
influencing, e.g., the populations dynamics, as well as the fluorescence spectrum.

Summarizing, the usual description of the amplification mechanism relies on the idea that the noise
acts on each constituent without being influenced by the presence of the others.
However, in general this cannot be taken for granted a priori.
Of course, this does not mean that a more realistic description of the $N$-particle system is incompatible with the amplification mechanism.
On the contrary, we think that such a description will help to clarify to what extent and how the action of the noise induces the localization
of the center of mass
on more complex systems.
This collective property of an $N$-particle system seems to be more motivated if the effects of the noise on the different
components are correlated.
Indeed, a limited, but consistent way to take into account the mutual interaction
between the constituents of the system is just obtained by 
dealing with a rigid body. 
The details of the internal dynamics are neglected, but each particle is considered as a component of a single $N$-particle bound system.
The action of the noise on each constituent 
is turned directly into the corresponding action on the center of mass: 
the overall effect of the noise is correctly described by the sum of the action
on each constituent, so that the results of the previous paragraph are in this case
justified.


\section{Conclusions}\label{sec:concl}

In this paper, we have extended the GRW model \cite{Ghirardi1986}
in order to avoid the infinite growth of the energy of the system.
We have introduced new jump operators, while leaving the other defining
features of the model unchanged. The jump operators correspond to the Fourier transform
of the Lindblad operators of a model for collisional decoherence \cite{Vacchini2000,Vacchini2001,Hornberger2006}
and depend
on the momentum operator of the system. Hence, they induce a dissipative evolution, which leads to
a damping of the momentum, as well as to a finite asymptotic value of the energy. The inclusion of a dissipative
mechanism within the collapse model has called for the definition of a new parameter,
which is directly linked to the temperature $T$ of the noise. The original GRW model is
retrieved in the high temperature limit $T \rightarrow \infty$.

We have proved that the collapse model can be formulated equivalently
by means of a proper stochastic differential equation, defined in terms of a random field made up of one counting process
for each point of space. 
The stochastic differential
equation fully determines the possible trajectories in the Hilbert space of the system.
By focusing on the case of gaussian initial states, we have investigated the main
features of these trajectories, showing in particular the occurrence of the position and momentum localization.
The latter clarifies that the energy divergence in the original GRW model
has to be traced back to fluctuations in momentum, rather than to an unlimited
increase of the momentum variance. 

As long as a rigid body is taken into account, the amplification mechanism still holds,
so that the center of mass of a macroscopic system is localized on very short
time scales and behaves for all the practical purposes according
to classical mechanics.
As usual, this has been shown by considering
a master equation for the center of mass which is given by the sum of the one-particle master equations and taking the partial
trace over the relative degrees of freedom. Nevertheless, we have also argued
that 
a more general and realistic analysis
of the $N$-particle system's evolution  should be given if
the interaction between the components of the system can affect the action of the noise.

Our results provide an important step toward the establishment
of a realistic jump collapse model, which generalizes the original GRW proposal.
Indeed, important advances are still to be made.
A promising possibility to recover entirely
the principle of the energy conservation
is to consider the noise as a real physical field influenced by the presence
of the system:
the energy variations of the latter could be then explained in terms of an energy exchange with the noise.
Let us emphasize how this approach could be put forward by associating the random field 
to the physical field, as proposed for the continuous-time  
collapse models \cite{Bassi2005,Bassi2013}.  
Furthermore, also in this paper we have assumed the hypothesis of a white noise: 
this greatly simplifies the characterization of the model and can provide
a satisfactory description of the interaction between the system and the noise in certain regimes.
However, there is no reason to exclude a priori the general case of a
colored noise, such that, e.g., the time distribution of the localization events is no longer memoryless.

Finally, the inclusion of dissipation has allowed us to avoid the energy divergence also in the CSL model \cite{Smirne2014},
by properly modifying the corresponding stochastic differential equation in the Fock space associated with the system,
and to investigate the overheating problem in the Di\'osi-Penrose model \cite{Diosi1989,Bahrami2014}.

\acknowledgments
The authors acknowledge financial support from the EU
project NANOQUESTFIT, INFN, the funding FRA-2014 of the University of Trieste and
COST (MP1006). They also thank G.C. Ghirardi and G. Gasbarri
for very useful discussions.

\appendix

\section{Localization mechanism for the superposition of two gaussian wavefunctions}\label{app:sovr}
In this Appendix we study the action of the jump operator $L_y(\widehat{X}, \widehat{P})$
on a single-particle state $\ket{\varphi}$ given by the superposition of two gaussian wavefunctions, with null mean momentum, equal variance and opposite mean positions, $\pm \xz$ ($\xz >0$),
see Eq.(\ref{eq:sovr}) and Fig. \ref{fig:1} {\bf(b)}, such that $\xz^2\gg \rc^2\gg \sig$.
First, note that the matrix elements of  $L_y(\widehat{X}, \widehat{P})$ in the position representation are given by
\begin{equation}\label{eq:braxL}
\bra{X} L_y(\widehat{X}, \widehat{P})\ket{X'} = (\pi  \rc^2(1+k)^2)^{-1/4} e^{-(X-y)^2/(2  \rc^2(1+k)^2)}\delta\left(\left(\frac{1-k}{1+k}\right)X-X' +\left(\frac{2k}{1+k}\right) y \right),
\end{equation}
where $k$ is defined in Eq.(\ref{eq:kg}). As long as $k \neq 0$, where $k=0$ corresponds to the original GRW model, the off-diagonal matrix elements
of the jump operator can be different from $0$.
As a consequence, one has that the state $\ket{\varphi_y}$ after the localization is, see Eq.(\ref{eq:jjd}),
\begin{eqnarray}
\scalar{X}{\varphi_y} &=& C_y \int \mathd X' \delta\left(\left(\frac{1-k}{1+k}\right)X-X' +\left(\frac{2k}{1+k}\right) y \right) 
e^{- (X-y)^2/(2 \rc^2(1+k)^2)}\left(c_+ e^{-(X'-\xz)^2/(2 \sig)} + c_- e^{-(X'+\xz)^2/(2 \sig)} \right), \nonumber\\
&=& C_y e^{- (X-y)^2/(2 \rc^2(1+k)^2)}\left(c_+ e^{-\left(\left(\frac{1-k}{1+k}\right)X +\left(\frac{2k}{1+k}\right) y-\xz\right)^2/(2 \sig)} + c_- e^{-\left(\left(\frac{1-k}{1+k}\right)X +\left(\frac{2k}{1+k}\right) y+\xz\right)^2/(2 \sig)} \right) \nonumber
\end{eqnarray}
compare with Eq.(\ref{eq:cmp}). Let's now focus on a localization process taking place at $y= \xz$,
so that
\begin{eqnarray}
\scalar{X}{\varphi_y} &=& C_y \left(c_+ e^{-\left((1-k)^2/(2\sig(1+k)^2) + 1/(2\rc^2(1+k)^2)\right)(X -\xz)^2} + c_- e^{- (X-\xz)^2/(2 \rc^2(1+k)^2)} e^{-\left(\left(\frac{1-k}{1+k}\right)X +\left(\frac{3k+1}{1+k}\right) \xz\right)^2/(2 \sig)} \right) \nonumber\\
&\approx & C_y \left(c_+ e^{-\left((1-k)^2/(2\sig(1+k)^2) + 1/(2\rc^2(1+k)^2)\right)(X -\xz)^2}  + c_- e^{- 2 \xz^2/( \rc^2 (1-k)^2)} e^{-\left(\left(\frac{1-k}{1+k}\right)X +\left(\frac{3k+1}{1+k}\right) \xz\right)^2/(2 \sig)} \right),
\end{eqnarray}
where we have used $\sig \ll \rc^2 (1-k)^2$. Still, the second term is strongly suppressed since $\xz^2 \gg \rc^2 (1-k)^2$: the localization mechanism leaves us
with only the gaussian wavefunction centered around $\xz$.

Let us now evaluate the probability that the localization position $y$ is such that $y+\xz \gg \rc $ (so that
it is far from $-\xz$). 
Proceeding exactly as before, one finds
$$
p(y) \approx  (\sqrt{\pi}  \rc (1-k))^{-1} |c_+|^2 e^{-(y-\xz)^2/(\rc^2(1-k)^2)},
$$
i.e. a gaussian centered around $\xz$ and with width $\tilde{\sig} =  \rc (1-k)/\sqrt{2}$. Then, if we consider the probability to
have a jump in a neighborhood of $\xz$, say, within $5 \tilde{\sig}$, we end up with $p(y) \approx (1- 6 \times 10^{-7}) |c_+|^2$:
on a proper coarse grained spatial scale (fixed by $ \rc(1-k)$) the predictions of quantum mechanics are recovered for all the practical purposes.

\section{Gaussian solutions of the stochastic differential equation}\label{sec:gs}
In this Appendix, we are going to study more in detail the trajectories described by Eq.(\ref{eq:sol}),
focusing on the case of gaussian wavefunctions.
We will show that the stochastic differential equation preserves the gaussian structure of the wavefunction,
whose evolution in time will be explicitly characterized. 

Let us take into account a gaussian initial state $\ket{\psi}_0 = \ket{\phi^{\xz,\pz,\sig}}$, see Eq.(\ref{eq:gaus}),
and assume $\widehat{H} = \widehat{P}^2/(2M)$, so that
the unitary evolution describes the usual spreading of the gaussian wave packet.
The transition from the initial state to the gaussian state immediately before the jump at time
$t_1$ can be summarized as
\begin{eqnarray}
 \xz &\longrightarrow& \xz_{t_1} = \xz + \frac{\pz}{M} (t_1-t_0)\nonumber\\
\pz  &\longrightarrow& \pz_{t_1} = \pz \nonumber \\
 \sig&\longrightarrow& \sig_{t_1} =\sig + \frac{i \hbar}{M} (t_1-t_0).    \label{eq:free}
\end{eqnarray}
The jump at time $t_1$ and position $y$ modifies the wavefunction according to Eq.(\ref{eq:xps}): explicitly
\begin{eqnarray}
\xz_{t_1} &\longrightarrow& \xzp_{t_1} =  g_{\sig_{t_1}} \xz_{t_1} +(1-g_{\sig_{t_1}}) y \nonumber\\
\pz_{t_1} &\longrightarrow&\pz'_{t_1} = \pz_{t_1} \frac{1 -k}{1+k} \nonumber\\
\sig_{t_1} &\longrightarrow&\sig'_{t_1} = \left(\frac{(1-k)^2}{\sig_{t_1} (1+k)^2}+\frac{1}{ \rc^2 (1+k)^2}\right)^{-1}, \label{eq:xpss}
\end{eqnarray}
where $g_{\sig_{t_1}}$ is defined as in Eq.(\ref{eq:kg2}). Hence, the wavefunction
has still a gaussian form after the jump, but since the free evolution implies a complex coefficient $\sig_{t_1}$,
both $\xzp_{t_1}$ and $\sig'_{t_1}$ will be in general complex numbers.
The full solution can be built up by iterating these two steps, where, of course, $t_0$ ($t_1$)
has to be replaced with the instant $t_{j-1}$ ($t_j$ ) of the $(j-1)$-th ($j$-th) jump
and $\xz, \pz, \sig$ in Eq.(\ref{eq:free}) have to be replaced with the parameters $\xzp_{t_{j-1}}, \pz'_{t_{j-1}}, \sig'_{t_{j-1}}$
after the $(j-1)$-th jump. Finally, after the last jump at time $t_m$, Eq.(\ref{eq:free}) has to be used once more,
with the instants $t$ and $t_m$, as well as the parameters  $\xzp_{t_m}, \pz'_{t_m}, \sig'_{t_m}$.
In conclusion, the solution $\ket{\psi(t)}$ is the gaussian wavefunction 
$ \ket{\phi^{\xz_t,\pz_t,\sig_t}}$  as in Eq.(\ref{eq:gaus}),
with $\xz_t, \sig_t \in \mathbbm{C}, \pz_t \in \mathbbm{R}$ and
\begin{equation}\label{eq:cc}
C = \left( \frac{\pi |\sig_t|^2}{\sig^{\text{R}}_t}\right)^{-1/4}e^{-\frac{(\xz^{\text{I}}_t)^2}{2\sig^{\text{R}}_t}- \frac{\pz_t \xz^{\text{I}}_t}{\hbar}}
\end{equation} 
the normalization constant, with $z^{\text{R}}$ and $z^{\text{I}}$, respectively, real and imaginary part of $z$;
moreover, $\sig>0$ implies  $\sig^{\text{R}}_{t}, \sig^{\text{I}}_{t},  g^{\text{R}}_{\sig_{t}}>0$,
while $g^{\text{I}}_{\sig_{t}}<0$. The parameters $\xz_t,\pz_t,\sig_t$ are stochastic quantities depending
on the specific trajectory $\omega_t$ (i.e. they are short-hand notation for $\xz(\omega_t),\pz(\omega_t),\sig(\omega_t)$).
In particular, $\pz_t$ only depends on the number of jumps up to time $t$: given a trajectory with $m$ jumps,
one has
\begin{equation}\label{eq:pzt}
\pz_t = \pz \left(\frac{1-k}{1+k}\right)^m.
\end{equation}
On the other hand, $\sig_t$ depends also on the instants at which the jumps take place, but not on 
their positions:
explicitly,
\begin{equation}\label{eq:fff}
\sig_t = \mathcal{G}_m\left(\mathcal{G}_{m-1}\left(\ldots \mathcal{G}_1(\gamma)\right)\right) +\frac{i \hbar \tau}{M},
\end{equation} 
with 
\begin{equation}\label{eq:f}
\mathcal{G}_j(x) =\left(\left(x+\frac{i \hbar \tau_j}{M}\right)^{-1}\left(\frac{1-k}{1+k}\right)^2+\frac{1}{\rc^2(1+k)^2}\right)^{-1},
\end{equation}
where $\tau_j \equiv t_j - t_{j-1}$ and $\tau \equiv t-t_m$.
Finally, $\xz_t$ depends also on the specific positions where the jumps take place: one has
\begin{eqnarray}\label{eq:xzt}
\xz_t &=&  \prod^m_{j=1}g_{\sig_{t_j}}\left(\xz + \frac{\pz \tau_1}{M}\right)+ \prod^m_{j=2}g_{\sig_{t_j}}\left((1-g_{\sig_{t_1}}) y_1 + \frac{\pz_{t_1}\tau_2}{M}\right) + \ldots \nonumber\\
&& +g_{\sig_{t_m}}\left((1-g_{\sig_{t_{m-1}}}) y_{m-1} + \frac{\pz_{t_{m-1}}\tau_{m}}{M}\right) + (1-g_{\sig_{t_{m}}}) y_m + \frac{\pz_{t_{m}}\tau}{M} .
\end{eqnarray}

For the sake of completeness, let us mention that the parameters $\xz_t,\pz_t,\sig_t$  
satisfy the following stochastic differential equations:
\begin{eqnarray}
\mathd \xz_t &=& \frac{\pz_t}{M} \mathd t  +  (1-g_{\sig_t})\int \mathd y\left(y - \xz_t \right) \mathd N_y(t)  \nonumber\\
\mathd \pz_t &=& - \frac{2k}{k+1} \pz_t \mathd N(t) \nonumber\\
\mathd \sig_t &=& \frac{i \hbar}{M} \mathd t + \left( \left(\frac{(1-k)^2}{\sig_t(1+k)^2}+\frac{1}{ \rc^2(1+k)^2}\right)^{-1}-\sig_t \right) \mathd N(t).
\end{eqnarray} 
The deterministic contributions describe the evolution of the parameters under the unitary part of the dynamics, while
the stochastic terms describe the action of the jumps.
The stochastic contribution in the equations of $\pz_t$ and $\sig_t$ traces back
to the Poisson process $N(t)$ defined in Eq.(\ref{eq:np}), which counts the overall occurrence
of the jumps, without any discrimination about their position.
 
From the evolution of the parameters
which fix the gaussian wavefunction, one can directly infer the evolution of relevant physical quantities.
Given a wavefunction $ \ket{\phi^{\xz_t,\pz_t,\sig_t}}$  as in Eq.(\ref{eq:gaus}),
with $\xz_t, \sig_t \in \mathbbm{C}, \pz_t \in \mathbbm{R}$, as well as the normalization $C$
in Eq.(\ref{eq:cc}), let us denote as $\langle O \rangle_t$ the mean value of the observable $\widehat{O}$ on such a state at time $t$, i.e.,
\begin{equation}\label{eq:avo}
\langle O \rangle_t  =  \bra{\phi^{\xz_t,\pz_t,\sig_t}} \widehat{O}\ket{ \phi^{\xz_t,\pz_t,\sig_t}},
\end{equation}
while we denote as $(\Delta_{\phi_t} O)^2$ the corresponding variance,
\begin{equation}\label{eq:apvar}
(\Delta_{\phi_t} O)^2  =  \bra{\phi^{\xz_t,\pz_t,\sig_t}} \widehat{O}^2\ket{ \phi^{\xz_t,\pz_t,\sig_t}}-(\bra{\phi^{\xz_t,\pz_t,\sig_t}} \widehat{O}\ket{ \phi^{\xz_t,\pz_t,\sig_t}})^2.
\end{equation}
The mean value of the position at time $t$  is
\begin{equation}\label{eq:meanx}
\langle X \rangle_t  = \xz^{\text{R}}_t + \frac{ \sig^{\text{I}}_t}{\sig^{\text{R}}_t}\xz^{\text{I}}_t,
\end{equation}
while the variance is 
\begin{equation}\label{eq:varx}
(\Delta_{\phi_t} X)^2 =\frac{|\sig_t|^2}{2 \sig^R_t} = \frac{1}{2 \om^R_t},
\end{equation}
with $\om_t = 1/\sig_t$.
Moreover, the mean value of the momentum at time $t$ is 
\begin{equation}\label{eq:momav}
\langle P \rangle_t  = \pz_t + \hbar \frac{\xz^{\text{I}}_t}{\sig^{\text{R}}_t},
\end{equation}
while the variance is 
\begin{equation}\label{eq:varp}
(\Delta_{\phi_t}P)^2  = \frac{\hbar^2}{2 \sig^{\text{R}}_t}.
\end{equation}
Hence, one has
\begin{equation}
(\Delta_{\phi_t} X)^2 (\Delta_{\phi_t} P)^2 = \frac{\hbar^2}{4}\left(1 + \left(\frac{\sig^{\text{I}}_t}{\sig^{\text{R}}_t}\right)^2\right),
\end{equation}
so that  $ \ket{\phi^{\xz_t,\pz_t,\sig_t}}$ will not be a minimum uncertainty state as long as $\sig^{\text{I}}_t \neq 0$.

Indeed, the quantities in Eqs.(\ref{eq:meanx})-(\ref{eq:varp}) do depend on the trajectory $\omega_t$: the variances
depend on the instants in which the jumps occur, while the mean values also depend on the positions
of the jumps, see Eqs.(\ref{eq:pzt})-(\ref{eq:xzt}). The explicit evaluation of these quantities then
calls for a numerical analysis, which can be easily achieved starting from the results of this Appendix.

\section{Asymptotic values of position and momentum variance for jumps equally spaced in time}\label{app:axp}

In order to estimate the asymptotic value of, respectively, position and momentum variance, it is convenient to
examine the trajectories in which all jumps are equally spaced in time \cite{Ghirardi1986},
so that $\tau_j = 1/ \lambda$ for any $j$ (where, of course, $\lambda$ has to be replaced with $\lambda_{\text{macro}}$
for a macroscopic system).

Since the position variance is fixed 
by $\sig_t$, see Eq.(\ref{eq:varx}),
a stable condition will be reached when a
cycle composed of a free evolution and the subsequent jump does not modify the value of $\sig$.
Explicitly, the equilibrium value $\sig_{\text{eq}}$ satisfies, see Eqs.(\ref{eq:fff}) and (\ref{eq:f}),
\begin{equation} 
\left(\left(\sig_{\text{eq}}+ i \epsilon\right)^{-1}\left(\frac{1-k}{1+k}\right)^2+\frac{1}{\rc^2(1+k)^2}\right)^{-1} = \sig_{\text{eq}},
\end{equation}
with $\epsilon = \hbar/(M \lambda)$.
This equation is solved by
\begin{equation}\label{eq:equ}
\sig_{\text{eq}} = \frac{1}{2 }\left( \jt- i \epsilon + \sqrt{(  \jt- i \epsilon)^2 + 4 i \epsilon \rc^2 (1+k)^2 } \right),
\end{equation}
where $\jt$ is the threshold value defined in Eq.(\ref{eq:soglia}).
Given a complex number $z = z^{\text{R}} + i z^{\text{I}}$, its square root can be expressed as
$$
\sqrt{z} = \zeta \sqrt{|z|}\left( \sqrt{\frac{1}{2}+\frac{|z^{\text{R}}|}{2 |z|}} + \text{sgn}( z^{\text{R}} z^{\text{I}}) i   \sqrt{\frac{1}{2}-\frac{|z^{\text{R}}|}{2 |z|}} \right),
$$
with  $\zeta=1$ if $z^{\text{R}} > 0$, $\zeta = i$ if $z^{\text{R}} < 0$ and $z^{\text{I}} \geq 0$, $\zeta = -i$ if $z^{\text{R}}, z^{\text{I}} < 0$
and $\text{sgn}(x)$ is the signum function.
By using Eq.(\ref{eq:equ}), we thus get Eq.(\ref{eq:asvar}) for the asymptotic value
of the position variance. Moreover, replacing Eq.(\ref{eq:equ}) in the expression
of the momentum variance in Eq.(\ref{eq:varp}), we get the asymptotic value in Eq.(\ref{eq:deltap}).

\section{Derivation of the equation for the characteristic function}\label{app:dote}
In this Appendix, we show that  Eq.(\ref{eq:megrwdiss}) for the statistical operator $\hat{\rho}(t)$
is equivalent to Eq.(\ref{eq:char}) for the characteristic function $\chi(\nu,\mu, t)$ defined
in Eq.(\ref{eq:chid}).

By Eq.(\ref{eq:megrwdiss}), we get
\begin{equation}
\partial_t \chi(\nu,\mu, t) = \mathcal{A}\left[\chi(\nu,\mu, t)\right] +  \mathcal{B}\left[\chi(\nu,\mu, t)\right] - \lambda  \chi(\nu,\mu,t). 
\end{equation}
Let us start with the Hamiltonian contribution:
\begin{eqnarray}
 \mathcal{A}\left[\chi(\nu,\mu, t)\right]  &=& -\frac{i}{\hbar}\left(\text{Tr}\left\{\widehat{H}\hat{\rho}(t)e^{\frac{i}{\hbar}\left(\nu \hat{X}+\mu \hat{P}\right)} \right\}  
 -\text{Tr}\left\{\hat{\rho}(t)\widehat{H}e^{\frac{i}{\hbar}\left(\nu \hat{X}+\mu \hat{P}\right)} \right\}  \right) = \frac{\nu}{M}\partial_{\mu} \chi(\nu,\mu,t),
\end{eqnarray}
where we used 
$$
\left[\widehat{P}^2 , e^{\frac{i}{\hbar}\left(\nu \hat{X}+\mu \hat{P}\right)} \right] = \nu \left( \widehat{P} ^{\frac{i}{\hbar}\left(\nu \hat{X}+\mu \hat{P}\right)}
+ e^{\frac{i}{\hbar}\left(\nu \hat{X}+\mu \hat{P}\right)}  \widehat{P} \right)
$$
and the cyclicity of the trace.

The dissipative term provides us with
\begin{eqnarray}
 \mathcal{B}\left[\chi(\nu,\mu, t)\right] &=&  \frac{\rc(1+k) \lambda}{\sqrt{\pi} \hbar}
\int \mathd \mom \, \text{Tr}\left\{ e^{\frac{i}{\hbar} \mom \widehat{X}} 
e^{-\frac{\rc^2}{2\hbar^2 }\left( (1+k) \mom+2 k \widehat{P}\right)^2} \hat{\rho}(t) 
e^{-\frac{\rc^2}{2\hbar^2 }\left( (1+k) \mom+ 2 k \widehat{P}\right)^2}e^{-\frac{i}{\hbar} \mom \widehat{X} }e^{\frac{i}{\hbar}\left(\nu \hat{X}+\mu \hat{P}\right)} \right\} \nonumber\\
&=& \frac{\rc(1+k) \lambda}{\sqrt{\pi} \hbar}
\int \mathd \mom \mathd X \, \bra{X}\left( e^{\frac{i}{\hbar} \mom \widehat{X}} 
e^{-\frac{\rc^2}{2\hbar^2 }\left( (1+k) \mom+2 k \widehat{P}\right)^2}\hat{\rho}(t) 
e^{-\frac{\rc^2}{2\hbar^2 }\left( (1+k) \mom+2 k \widehat{P}\right)^2} e^{-\frac{i}{\hbar} \mom \widehat{X} }e^{\frac{i}{\hbar} \mu \hat{P}}
e^{\frac{i}{\hbar} \nu \hat{X}}\right)\ket{X} e^{-\frac{i}{2} \frac{\mu \nu}{\hbar}} \nonumber\\
&=& \frac{\rc(1+k) \lambda}{\sqrt{\pi} \hbar}
\int \mathd \mom \mathd X \mathd X' \mathd X'' \,  e^{\frac{i}{\hbar} \mom \mu} 
f(X, X')\rho(X', X'', t) 
f(X'', X-\mu) 
e^{\frac{i}{\hbar} \nu X} e^{-\frac{i}{2} \frac{\mu \nu}{\hbar}},\label{eq:prev}
\end{eqnarray}
where we used $e^{\frac{i}{\hbar} \mu \hat{P}} \ket{X} = \ket{X-\mu}$ and we have introduced
\begin{eqnarray}
f (X,X') &=& \bra{X} e^{-\frac{\rc^2}{2\hbar^2 }\left( (1+k) \mom+2 k \widehat{P}\right)^2} \ket{X'} = \int \frac{\mathd P}{2 \pi \hbar} e^{\frac{i}{\hbar} P (X-X')}
e^{-\frac{\rc^2}{2\hbar^2 }\left( (1+k) \mom+2 k \widehat{P}\right)^2} = \frac{1}{2 \sqrt{2 \pi} k \rc} e^{- \frac{1}{8 k^2 \rc^2}(X-X')^2} e^{-\frac{i(1+k) \mom}{2\hbar k}(X-X')}\nonumber.
\end{eqnarray}
Inserting this relation and Eq.(\ref{eq:inv}) in Eq.(\ref{eq:prev}), we obtain
\begin{eqnarray}
 \mathcal{B}\left[\chi(\nu,\mu, t)\right]  
&=& \frac{\lambda(1+k)}{16 \pi^{5/2} \hbar^2 k^2 \rc}
\int \mathd \mom \mathd X \mathd X' \mathd X''  \mathd \nu' \, e^{\frac{i}{\hbar} (\mom-\nu/2) \mu} e^{- \frac{i}{2 \hbar} \nu' (X'+X'')} \chi(\nu', X'-X'',t)
 \nonumber\\
&&\times e^{- \frac{1}{8 k^2 \rc^2}\left[(X-X')^2+(X''-X + \mu)^2\right]} e^{-\frac{i(1+k) \mom}{2\hbar k}(X''-X' + \mu)}
e^{\frac{i}{\hbar} \nu X} \nonumber\\
&=& \frac{ \lambda(1+k)}{8 \pi^{2} \hbar^2 k}
\int \mathd \mom  \mathd X' \mathd X''  \mathd \nu' \, e^{\frac{i}{\hbar} \mom \mu} e^{- \frac{i}{2 \hbar} \nu' (X'+X'')} \chi(\nu', X'-X'',t) \nonumber\\
&&\times e^{-\frac{i(1+k) \mom}{2 \hbar k}(X''-X' + \mu)}
e^{-\frac{\nu^2 \rc^2 k^2}{\hbar^2}} e^{-\frac{1}{16 k^2 \rc^2}(X'-X''-\mu)^2}e^{i\frac{\nu}{2\hbar}(X'+X'')}\nonumber\\
&=&\frac{ \lambda(1+k)}{16 \pi^{2} \hbar^2 k}
\int \mathd \mom  \mathd s \mathd \Delta  \mathd \nu' \, e^{\frac{i}{\hbar} \mom \mu} e^{- \frac{i}{2 \hbar} (\nu'-\nu) s} \chi(\nu', \Delta,t) e^{\frac{i(1+k) \mom}{2\hbar k}(\Delta - \mu)}
e^{-\frac{\nu^2 \rc^2 k^2}{\hbar^2}} e^{-\frac{1}{16 k^2 \rc^2}(\Delta-\mu)^2}\nonumber\\
&=&\frac{ \lambda(1+k)}{4 \pi \hbar k}
\int \mathd \mom \mathd \Delta  \, e^{\frac{i}{\hbar} \mom \mu} \chi(\nu, \Delta,t) e^{\frac{i(1+k) \mom}{2 \hbar k }(\Delta - \mu)}
e^{-\frac{\nu^2 \rc^2 k^2}{\hbar^2}} e^{-\frac{1}{16 k^2 \rc^2}(\Delta-\mu)^2}\nonumber\\
&=& \lambda
 \chi\left(\nu, \mu\left(\frac{1-k}{1+k}\right),t\right) 
e^{-\frac{\nu^2 \rc^2 k^2}{\hbar^2}} e^{-\frac{\mu^2}{4 \rc^2(1+k)^2}}.
\end{eqnarray}
This concludes the proof.


\begin{thebibliography}{10}
\bibitem{note2}
As a paradigmatic example consider that the stochastic average of the pure states $\ket{\psi(\omega_t)}\bra{\psi(\omega_t)}$ will generally be a mixed state $\hat{\rho}(t)$.
\bibitem{nota3}
Indeed, this is the case, more in general, for the time-evolution of any operator $f(\widehat{P})$ if one moves to the Heisenberg picture.
\end{thebibliography}
\end{document}